%% file: acl_latex.tex
\pdfoutput=1
\documentclass[11pt]{article}

\usepackage[final]{acl}

\usepackage{times}
\usepackage{latexsym}

\usepackage[T1]{fontenc}
\usepackage[utf8]{inputenc}

\usepackage{microtype}

\usepackage{inconsolata}

\usepackage{graphicx}

\usepackage{subcaption}
\usepackage{enumitem}
\usepackage{multicol}
\usepackage{tikz}% tikz
\usepackage{amsmath}
\usepackage{amssymb}
\usepackage{mathtools}
\usepackage{amsthm}
\usepackage{multirow}
\usepackage[capitalize,noabbrev]{cleveref}

\usepackage{wrapfig}
\usepackage{makecell}
\usepackage{booktabs}       % professional-quality tables
\usepackage{xspace}
\usepackage{authblk}

\newcommand{\mypara}[1]{\noindent{\bf {#1}.}~}
\newcommand{\ie}[0]{{i.e.,}\xspace}
\newcommand{\eg}[0]{{e.g.,}\xspace}

\newenvironment{packeditemize}{
\begin{list}{$\bullet$}{
\setlength{\itemsep}{0pt}
\addtolength{\labelwidth}{10pt}
\setlength{\leftmargin}{12pt}
\setlength{\listparindent}{\parindent}
\setlength{\parsep}{2pt}
\setlength{\topsep}{0pt}}}
{\end{list}
}
\newcolumntype{Y}{>{\centering\arraybackslash}X}

\def\Snospace~{\S{}}

\title{Large Language Models as Realistic Microservice Trace Generators}

\author{
    \bf{Donghyun Kim$^{1}$,\ \ Sriram Ravula$^{1}$,\ \ Taemin Ha$^{1}$,} \\
    \bf{Alexandros G. Dimakis$^{2,3}$,\ \ Daehyeok Kim$^{1}$,\ \ Aditya Akella$^{1}$} \\ \\
  \texttt{\normalsize \{donghyun, sriram.ravula, taemin.ha\}@utexas.edu,}\\
  \texttt{\normalsize alexdimakis@berkeley.edu, \{daehyeok,akella\}@cs.utexas.edu} \\ \\
  {$^1$ University of Texas at Austin} 
  {$^2$ University of California, Berkeley} 
  {$^3$ BespokeLabs.ai}
}

\newcommand{\sys}{\texttt{TraceLLM}}

\begin{document}
\maketitle
\begin{abstract}
Workload traces are essential to understand complex computer systems' behavior and manage processing and memory resources.
Since real-world traces are hard to obtain, synthetic trace generation is a promising alternative.
This paper proposes a first-of-a-kind approach that relies on training a large language model (LLM) to generate synthetic workload traces, specifically {\em microservice call graphs}. 
To capture complex and arbitrary hierarchical structures and implicit constraints in such traces, we propose to train LLMs to generate recursively, making call graph generation a sequence of more manageable steps.
To further enforce learning constraints on the traces and generate uncommon situations, we apply additional instruction tuning steps to align our model with the desired trace features.
With this method, we train \sys{}, an LLM for microservice trace generation, and demonstrate that it produces diverse, realistic traces under varied conditions, outperforming existing approaches in both accuracy and validity.
The synthetically generated traces can effectively replace real data to optimize important microservice management tasks.
Additionally, \sys{} adapts to downstream trace-related tasks, such as predicting key trace features and infilling missing data.
\end{abstract}

\input{introduction}
\input{background}
\input{method}
\input{eval}
\input{conclusion}

\bibliography{acl_latex}

\appendix
\include{appendix}

\end{document}

%% file: introduction.tex
\section{Introduction}\label{sec:intro}

Computer system workload traces document hardware or software events that occur as applications execute on computing machines, receive requests, process them, and serve responses. Such traces are vital for analyzing complex computer systems and optimizing their CPU, memory, networking resource allocation, and management.
However, obtaining access to real-world traces is often challenging due to limited public data availability and the difficulty of collecting them at large scale from diverse environments, especially in complex cloud computing settings.
As an alternative, synthetic traces provide limitless size and variety, offering significant advantages for testing and analysis, including the ability to simulate challenging conditions like stress-testing environments. While recent advances in generative machine learning, including LSTMs \citep{rnn}, GANs \citep{gan}, and diffusion models \citep{diffusion}, have improved synthetic trace generation, these methods typically only generate specific fields, such as the number of requests or resource types \citep{vm_trace}, or are confined to fixed-structure traces, like network packets~\citep{netfusion, netshare}.

In this paper, we propose \sys{} that adapts pre-trained large language models (LLMs)~\citep{gpt3,touvron2023llama} to generate synthetic workload traces.
LLMs have been successfully adapted beyond natural language to domains like protein sequences~\citep{tag-llm}, code~\citep{roziere2023code}, and tabular data~\citep{borisov2023language}. 
In addition, LLMs can produce outputs well-aligned with user inputs through fine-tuning~\citep{Ouyang2022TrainingLM, wei2021finetuned} and generalize to new prompts at inference~\citep{flan, t_zero}.
Thus, we posit that LLMs have the potential to generate synthetic traces that accurately capture real-world systems trace structures while adhering to user specifications. 

Despite their potential, using LLMs for synthetic systems trace generation presents significant challenges.
Traces are often logged in tabular format and structured as graphs, which can vary in depth and width.
Representing these traces as text sequences, optimal for modern autoregressive LLMs, is non-trivial.
Moreover, trace data often contain complex implicit constraints that rely on relationships between multiple trace features. 
For example, an application process's start time must precede that of all its child processes, while the parent process's end time must be later than the end time of its children; such constraints need to hold across all nodes in the application's graphical representation.

In this paper, we design \sys{}  to address these challenges, with a particular focus on generating \emph{microservice call graphs}. Microservices are the de facto approach to designing modern cloud-based applications and power many popular services today~\citep{uber,netflix,amazon}. Microservice call graphs trace the execution of user requests through such applications. These call graphs form a type of trace with a rich directed acyclic graph (DAG) structure. \sys{} represents these graphs in a text-based format suitable for LLMs, enabling their use in trace generation. Crucially, one of our key innovations is to generate call graphs with structural constraints by \emph{recursively generating subgraphs}, or {layers}. This approach allows the model to break down the complex task of reasoning about hierarchical graph structures and complex constraints into multiple simpler tasks. To further enhance the model's ability to follow structural constraints and to meet user-requested attributes, we employ instruction tuning. During this phase, the model learns to explicitly generate a series of {\em intermediate instructions}  between recursive layer generation steps, performing arithmetic and logical checks to ensure strict adherence to the desired structure.

We implement \sys{} by training Llama-2 7B~\citep{touvron2023llama} on microservice trace data and evaluate it through extensive evaluations. Results show that recursive generation with intermediate instructions significantly improves the model's ability to generate valid outputs for complex call graphs. Compared to generative and probabilistic models, our synthetic traces better align with real trace distributions. Furthermore, we demonstrate that synthetic traces effectively replace real data for training microservice management tasks and that our fine-tuned model excels Llama-3.1 405B~\citep{dubey2024llama} in trace feature prediction.
We release our codes in \href{https://github.com/ldos-project/TraceLLM}{\texttt{https://github.com/ldos-project/TraceLLM}}.

%% file: background.tex
\section{Background}
\label{sec:background}
\label{sec:trace_background}

\begin{figure*}
\centering
  \includegraphics[width=0.9\linewidth]{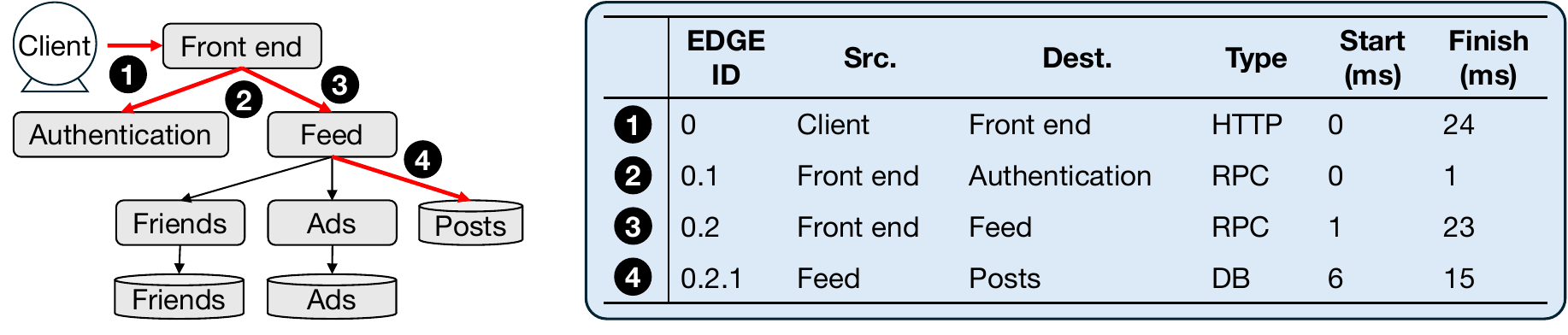}
  \vspace{-0.2cm}
  \caption{\small A simple social network application consists of eight microservices~\citep{meta_microservice}.
  Each user request triggers a sequence of microservice calls, forming a microservice call graph.
  The red lines represent the microservice call graph for a user request.
  Microservice call graphs are commonly logged in a tabular format, as shown in the figure on the right. Each row in the table represents a communication between two microservices with features in columns.}
  \label{fig:background_microservice}
  \vspace{-0.5cm}
\end{figure*}

\mypara{Microservice Call Graphs} In modern software architecture, an application is typically constructed as a constellation of multiple microservices~\citep{deathstarbench, madu, meta_microservice}, each with specific functionalities and dependencies on one another.
When a user interacts with an application, such as sending an HTTP request, a complex sequence of communications among these microservices is triggered. Thus, a user request induces a microservice \emph{call graph}, which maps the control/data flow and dependencies among the microservices involved in fulfilling the user's request.

\autoref{fig:background_microservice} illustrates a social network application with eight microservices.
Red arrows indicate communications between microservices involved in processing the user's request.
The request first reaches a microservice (e.g., “Front end” in~\autoref{fig:background_microservice}) and waits for the communication to terminate. If the microservice requires additional communication to handle the request, then it triggers another microservice call (e.g., from “Front end” to “Authentication” in~\autoref{fig:background_microservice}).
The communications triggered by a user's request form a microservice call graph with four microservices.
The vertices of the graph correspond to microservices (or the client), while the edges correspond to API calls invoking the microservices.
Note that not all edges appear in the graph, as some services may not be invoked for a given request.

A call graph can be represented as a tabular trace capturing API call features~(\ie edges) such as request source/destination, request type (\eg HTTP and RPC), and start/finish times. Given their \emph{hierarchical structure}, the tabular trace should preserve the parent-child relationships by ensuring that the child's source matches the parent's destination. Also, the start/end times of each call should be consistent with each other: (1) a microservice’s start time must precede its finish time, and (2) the parent-child relationships must be honored, i.e., the parent's start (finish) time must precede (follow) the child's. Finally, the IDs within a call graph (dot-decimal numbers provided for each call) must also be hierarchically connected to form a DAG.

\mypara{Synthetic Trace Generation using Machine Learning}
Microservice traces play a pivotal role in designing and evaluating techniques for improving the performance and reliability of microservice applications and optimizing the use of underlying resources.
Representative use cases include techniques for critical path analysis~\citep{crisp}, anomaly detection~\citep{tracevae},
root cause analysis~\citep{rcd}, cluster management~\citep{firm}, and cluster scheduling~\citep{atoll}.
Unfortunately, obtaining diverse real-world traces to study such techniques thoroughly remains challenging as publicly available traces are typically limited in size and only cover specific narrow settings far from the diversity expected when operating in the cloud.

Given the importance and scarcity of public computer system traces, including microservice traces, recent studies have explored generative models for synthetic trace generation. Existing works \citep{doppelganger, netfusion} leverage GAN~\citep{gan} and diffusion~\citep{diffusion} models to generate network packet traces, while other work~\citep{vm_trace} uses LSTMs~\citep{rnn} to generate virtual machine workload traces.
While these generative models are effective in their domains, the methods are limited to predicting specific fields or following training data distributions without conforming to structural constraints. These methods do not apply to microservice call graphs which requires handling hierarchical structures.

Since traces are structured and can be represented in tabular form, machine learning methods for synthetic tabular data generation could be applied to synthetic trace generation. Recent approaches, such as TVAE~\citep{ctgan} and GReaT~\citep{borisov2023language}, leverage VAE~\citep{vae} and language models to advance synthetic tabular data generation techniques. However, these methods do not account for the hierarchical structure of call graphs within tabular representations. We provide a detailed comparison in~\autoref{sec:eval}.

%% file: method.tex
\section{Training LLMs for Microservice Traces}
\label{sec:method}

Our goal is to train \sys{}, an LLM for microservice call graph traces, enabling end-users to simulate diverse scenarios, such as rare microservice invocations that exhibit high response times.
To achieve this, we condition the model's output on user-specified attributes, including the invoked application, the number of microservice communications (\ie graph edges), and overall application latency. Given the limitations of existing trace generation methods, we leverage LLMs. We initialize our model from LLMs trained on large text datasets, as these models have proven effective when adapted to specialized domains such as proteins~\citep{tag-llm} and code~\citep{roziere2023code}. Moreover, LLMs support flexible conditioning mechanisms, including natural language prompts~\citep{Ouyang2022TrainingLM} and structured input sequences~\citep{borisov2023language}.

This section presents our two-stage approach for training \sys{} to generate microservice call graphs. First, we extend LLMs with additional pre-training on a large corpus of microservice call graph traces, allowing the model to capture interaction patterns in real-world call graphs.
We also introduce recursive subgraph generation to improve its ability to produce large, structured graphs.
Second, we instruction fine-tune the model, enabling it to generate call graphs with user-specified attributes and ensuring constraint adherence through natural language reasoning.

\subsection{Pre-training}
\label{subsec:pre-training}

\begin{figure*}
    \centering
    \includegraphics[width=0.9\linewidth]{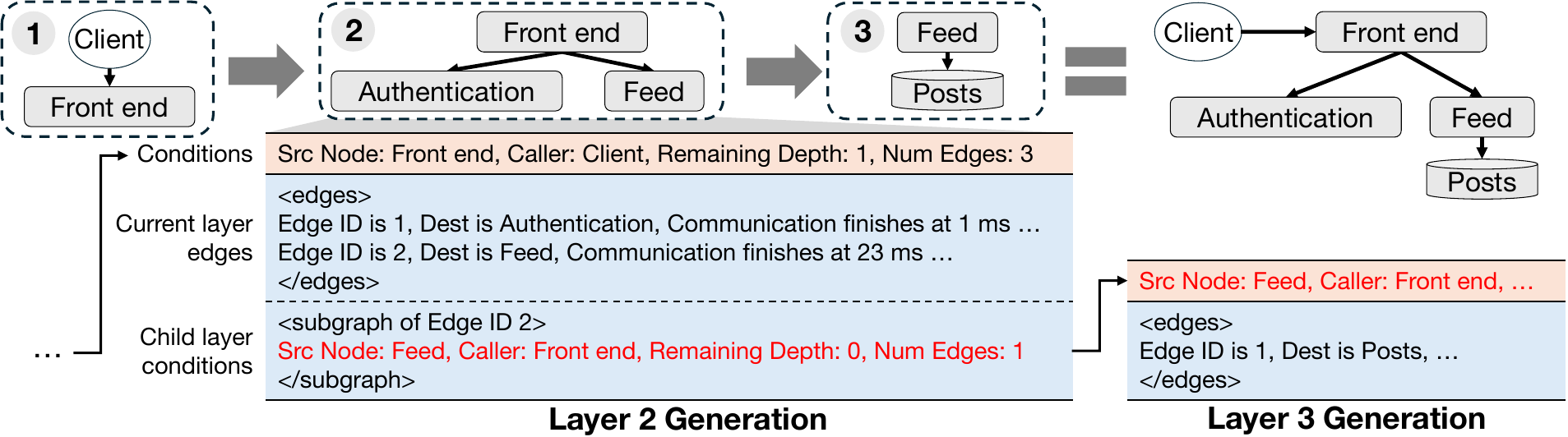}
    \vspace{-0.2cm}    
    \caption{\small Overview of the recursive generation method with a simplified example. 
    The model uses conditions generated in \textit{Layer 1} (\eg source node, caller, number of edges) to generate two edges in \textit{Layer 2}, one leading to \textit{Authentication} and the other to \textit{Feed}. The model also generates starting conditions for the next layer, beginning from the \textit{Feed} microservice. This recursion continues until all edges in \textit{Layer 3} are generated.
    }
    \label{fig:recursive_gen}
    \vspace{-0.5cm}
\end{figure*}

We pre-train our model on call graphs using an autoregressive language modeling objective. This stage adapts the general-purpose LLM, which was previously trained to model natural language text sequences, to the more specialized domain of microservice call graphs.

\subsubsection{Encoding Call Graphs as Text}
\label{subsec:encoding_call_graphs}
LLMs expect sequences of text as input, so we must encode our dataset of call graphs into text-based representations before training our model. As detailed in \autoref{sec:trace_background} and shown in \autoref{fig:background_microservice}, microservice call graphs are initially stored as tables. Rows represent edges (i.e., communications between microservices), while columns describe features for each edge. We follow the method proposed by GReaT~\citep{borisov2023language} and encode features in a natural language format. Our encoding procedure preserves all necessary information to recover the unique graph that produced the tabular data. Besides edge features, we also encode global attributes of the call graph to serve as conditioning information for the model.

A tabular call graph \( \mathbf{X} \) has \( m \) feature columns \(\{f_1, \dots, f_m\}\) and \( n \) edge rows \(\{\mathbf{x}_1, \dots, \mathbf{x}_n\}\), where the value of feature \( j \) for edge \( i \) is \( v_{ij} \). Each edge \( \mathbf{x}_i \) is encoded as a text sequence \( \mathbf{t}_i = [t_{i1}, \dots, t_{im}] \), where \( t_{ij} = [\phi(f_j), v_{ij}] \). Here, \( \phi(f) \) converts feature name \( f \) into a natural language template describing \( v_{ij} \). For example, the encoding for edge 1 in \autoref{fig:background_microservice} would be: [\texttt{Edge ID is 0, Source is Client, Destination is Front end, Type is HTTP, Communication starts at 0 ms, Communication finishes at 24 ms}]. The full graph is represented as \( \mathbf{t} = [\mathbf{t}_1, \dots, \mathbf{t}_n] \), a sequence of text-encoded edges. Since call graph structure depends on feature values and not column order, we randomly shuffle feature order within each edge during training \cite{borisov2023language} to eliminate spurious position-based associations.

The overall call graph can be described by attributes such as maximum depth, total edges, and total communication latency. These attributes summarize complex interactions and serve as prompts for call graph generation. Let call graph \( \mathbf{X} \) have \( r \) attributes with names \( \{a_1, \dots, a_r\} \) and values \( \{w_1, \dots, w_r\} \). We encode them as a text sequence \( \mathbf{c} = [c_1, \dots, c_r] \), where \( c_j = [a_j, ``:", w_j] \). See the \textit{Conditions} in red in \autoref{fig:recursive_gen} for an example. Attributes are prepended to each text-encoded call graph and predicted alongside edges during pre-training. Like edge features, attributes are randomly shuffled, and each is independently dropped with probability \( p_{drop} \) to enable flexible prompting.

\subsubsection{Recursive Generation}
\label{sec:recursive}

We propose to break down the task of generating a call graph into a series of recursive \textit{layer generation} tasks to handle complex structures. Starting from the initial attributes, or \textit{prompt} $\mathbf{c}$, the task for the model at each layer is to generate the edges originating from the \textit{Start Node} specified in the prompt. The model also generates a new prompt for the next layer based on the previous layer prompt and the edges generated in the current layer. This new prompt is then re-used to condition the model's output for the next layer. The recursive process continues until the requested attributes $\mathbf{c}$ are met.

Formally, for an encoded call graph $\mathbf{t} = [\mathbf{t}_1, \mathbf{t}_2, \dots, \mathbf{t}_n]$, we partition the edges $\mathbf{t}_i$ into a sequence of layers $[\mathbf{t}^1, \mathbf{t}^2, \dots, \mathbf{t}^l]$, where $l \leq n$. Each layer consists of a sequence of edges that share the same parent (\ie source) node, ensuring that no edge appears in multiple layers. For call graph conditions $\mathbf{c}$ that describe $\mathbf{t}$, we introduce layer conditions $\mathbf{c}^j$, $j \in \{1,2,\dots,l+1\}$. Layer condition $\mathbf{c}^j$ encodes the attributes of the remaining portion of the call graph after the sequence of layers $[\mathbf{t}^1, \mathbf{t}^2, \dots, \mathbf{t}^{j-1}]$ has been generated, and we define $\mathbf{c}^1 \coloneqq \mathbf{c}$ and $\mathbf{c}^{l+1} \coloneqq \emptyset$. We decompose the conditional call graph distribution as a chain of conditional layer distributions:
$p(\mathbf{t}|\mathbf{c}) = \prod_{k=1}^{l} p(\mathbf{c}^{k+1}, \mathbf{t}^k | \mathbf{c}^{k})$.
In other words, the model predicts call graphs from user prompts iteratively layer-by-layer. For layer $k$ the model takes $\mathbf{c}^{k}$ as input and produces the sequence of edges $\mathbf{t}^{k}$ followed by the conditions $\mathbf{c}^{k+1}$ of the next layer.
The process continues recursively, using $\mathbf{c}^{k+1}$ to predict the next layer, $k+1$. \autoref{fig:recursive_gen} illustrates an example of this recursive generation.

\subsection{Instruction Tuning}
\label{subsec:instruction_tuning}

After pre-training, we perform supervised fine-tuning to enhance the model’s ability to generate call graphs based on user instructions. Unlike pre-training, we exclude the initial call graph attributes \( \mathbf{c} \) (equivalent to the first-layer conditions \( \mathbf{c}^1 \)) from the loss computation, treating them as a fixed prompt. Users can provide additional instructions, and \autoref{subsec:instruction} presents results for two instruction types. To further aid reasoning, we supplement instructions with programmatically generated prompts that convert numerical and non-linguistic attributes (\eg application IDs) into natural language, as detailed in \autoref{appendix:training_data_example}.

\subsubsection{Intermediate Instructions}
\label{subsec:intermediate}

The model often struggles to generate consistent next-layer conditions \( \mathbf{c}^{k+1} \) based on the current layer’s edges \( \mathbf{t}^k \) and conditions \( \mathbf{c}^k \), sometimes violating physical constraints (e.g., assigning a layer higher latency than the overall call graph). Inspired by work showing that LLMs improve with explicit step-by-step reasoning \citep{wei2022chain, 51142}, we introduce natural language reasoning steps to reinforce constraint adherence. For example, we (1) compute remaining edges from the \textit{Num Edges} attribute in \( \mathbf{c}^k \) and edges in \( \mathbf{t}^k \), and (2) derive the \textit{Remaining Depth} in \( \mathbf{c}^{k+1} \) as \texttt{Child's remaining depth = current remaining depth - 1 = ...}. These \textit{intermediate instructions} are inserted before \( \mathbf{c}^{k+1} \) during instruction fine-tuning. We give an example of these steps in~\autoref{appendix:training_data_example}. 

%% file: eval.tex
\section{Evaluation} \label{sec:eval}
We thoroughly demonstrate the effectiveness of \sys{} in two major aspects: (1) synthetic trace quality in terms of structural validity~(\autoref{subsec:structured_reasoning}), distribution similarity~(\autoref{subsec:distribution_similarity}), and usefulness to train and evaluate machine learning-driven microservice management tasks~(\autoref{subsec:use_cases}), and (2) benefits from our use of LLMs in terms of instruction-following capabilities~(\autoref{subsec:instruction}) and trace-related downstream task performance~(\autoref{subsec:downstream_tasks}).

We initialize our model from Llama-2 7B~\citep{touvron2023llama} and train with LoRA~\citep{hu2022lora} on 1.36 million microservice call graph samples from the Alibaba v2022 dataset~\citep{madu}, corresponding to 1.1B tokens.
Further details on data preprocessing and training hyperparameters are provided in \autoref{appendix:training_details}.
We compare synthetic trace quality with various structured data generation methods such as GReaT~\citep{borisov2023language} and TVAE~\citep{ctgan}, and downstream task performance with one of the state-of-the-art LLMs, Llama-3.1 405B~\citep{dubey2024llama}.

\begin{figure*}
    \centering
    \begin{subfigure}{.45\textwidth}
        \includegraphics[width=\linewidth]{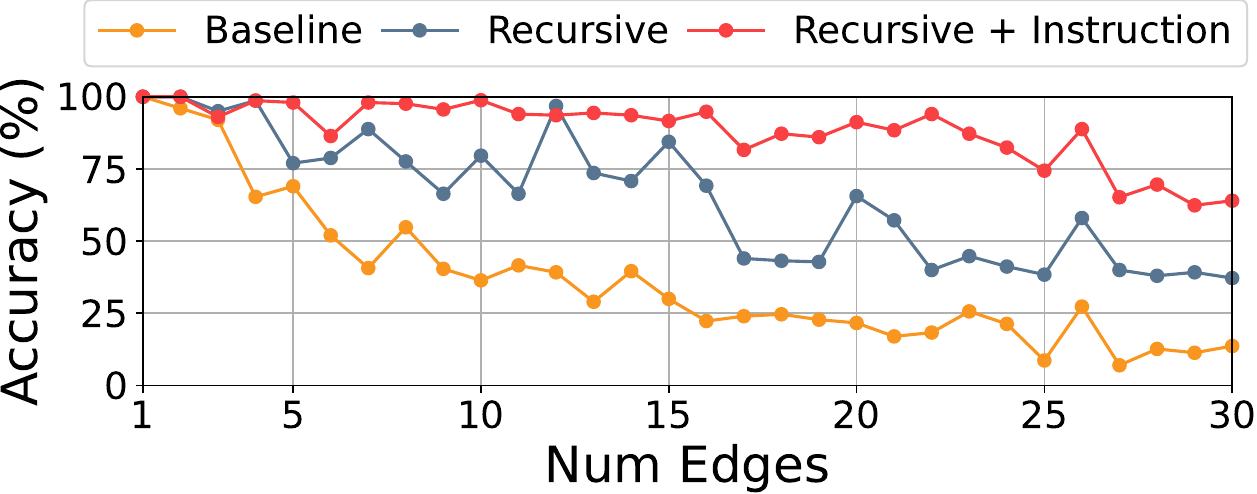}
        \caption{accuracy vs. number of edges}
        \label{fig:accuracy_by_edges}
    \end{subfigure}
    \begin{subfigure}{.25\textwidth}
        \includegraphics[width=\linewidth]{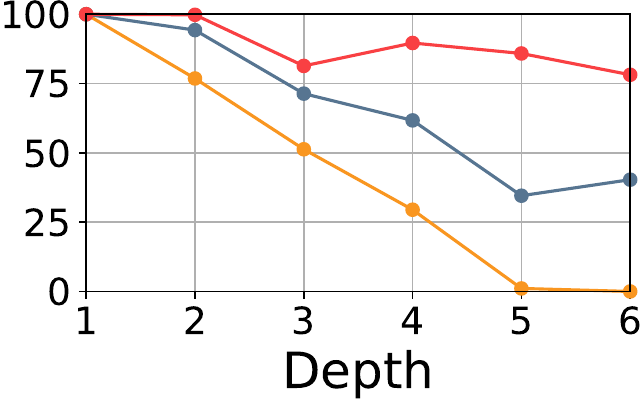}
        \caption{accuracy vs. depth}
        \label{fig:accuracy_by_depth}
    \end{subfigure}
    \begin{subfigure}{.25\textwidth}
        \includegraphics[width=\linewidth]{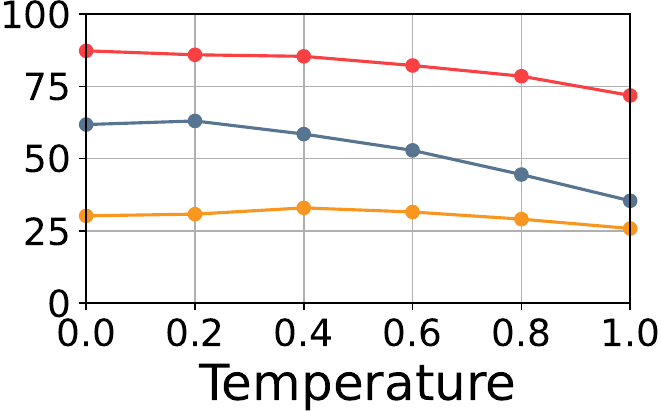}
        \caption{accuracy vs. temperature}
        \label{fig:accuracy_by_temp}
     \end{subfigure}
    \caption{\small Call graph generation accuracy with varying (a) edges and (b) depth in prompt using greedy sampling. (c) shows the accuracy with varying sampling temperature.  Accuracy measures the fraction of generated traces that are valid and follow the initial instructions. As shown, both recursive generation and instruction tuning improve the accuracy of the synthetic traces.}
    \label{fig:accuracy}
    \vspace{-0.3cm}
  \end{figure*}
  
\subsection{Structured Reasoning Accuracy} \label{subsec:structured_reasoning}
This experiment demonstrates how recursive generation and instruction tuning with intermediate instructions enhance LLMs' ability to accurately construct microservice call graphs. We evaluate our model by generating traces with specified \texttt{num\_edges} and \texttt{depth}.
A trace is deemed accurate if it correctly matches the specified \texttt{num\_edges} and \texttt{depth} and adheres to all structural constraints, such as valid DAG formations and appropriate start/finish times for communications, detailed in \autoref{appendix:validation_rules}. 
We generate 50 samples for each (\texttt{num\_edges}, \texttt{depth}) pair across ranges of $1 \leq$ \texttt{num\_edges} $\leq 30$ and $1 \leq$ \texttt{depth} $\leq 6$.

\mypara{Baselines}
We compare our model (\textit{recursive + instruction}) to Llama-2 7B models trained on text-encoded call graphs (1) without recursive generation and tuning with intermediate instructions (\textit{baseline}) and (2) with recursive generation but no instruction tuning (\textit{recursive}). Both baseline models are given \texttt{num\_edges} and \texttt{depth} as inputs during training (see \autoref{fig:baseline_training_example} for a training data example of the \textit{baseline} model). Baselines are trained using the same hyperparameters and number of tokens as our model.
The \textit{baseline} model follows GReaT~\citep{borisov2023language}, representing call graph traces as the tabular data format.

\mypara{Results}
\autoref{fig:accuracy_by_edges} and \autoref{fig:accuracy_by_depth} present the accuracy of generated call graphs across varying numbers of edges and depths. Generally, as complexity increases (\ie more edges or greater depth), the baseline model's accuracy decreases significantly—dropping below 25\% for edges $>$ 15 and nearing zero for depths $>$ 4. In contrast, the recursive generation model maintains higher accuracies, by approximately 30\% and 35\%, respectively. This improved performance is attributed to the model breaking down complex generation tasks into simpler, more manageable sub-tasks.

\autoref{fig:accuracy_by_temp} illustrates how decoding temperature affects accuracy. 
Both models show decreased performance as the temperature increases, but the recursive model consistently outperforms the baseline, maintaining about 10\% higher accuracy even at a temperature of 1.
Further, instruction tuning enhances model accuracy by 23\% to 36\% by directing the model to adhere to specific generation instructions, such as the number of edges per layer, which are outlined in~\autoref{appendix:training_data_example}. 

More results on accuracy with varying model sizes and memorization are in~\autoref{appendix:accuracy_varying model size} and~\autoref{appendix:memorization}.

\subsection{Similarity of Real and Synthetic Traces} \label{subsec:distribution_similarity}
To assess the quality of synthetic traces, we compare them to real traces from the validation dataset. We generate 50K call graphs using prompts derived from the validation set and measure their similarity to the original traces.

\mypara{Baselines}
We compare the following synthetic trace generation methods: 
\begin{packeditemize}
    \item \textbf{GReaT~\citep{borisov2023language} (Llama-2 7B + tabular format)}: A Llama-2 7B model fine-tuned on the tabular data format of call graph traces (Same as \textit{baseline} in~\autoref{subsec:structured_reasoning}).

    \item \textbf{Probabilistic model}: A call graph generator from Alibaba~\citep{alibaba_microservice_v2021} that samples graph structures based on statistical distributions, such as communication types and the number of child nodes per depth.
    
    \item \textbf{TVAE~\citep{ctgan}}: A VAE-based tabular data generator~\citep{gan}. Since it cannot directly generate traces, we use it to compare edge distributions. Training is limited to 100K randomly selected samples.
    
\end{packeditemize}

\begin{figure*}[!tbp]
\begin{minipage}[]{0.68\textwidth}
    \begin{subfigure}{.6\textwidth}
        \includegraphics[width=\linewidth]{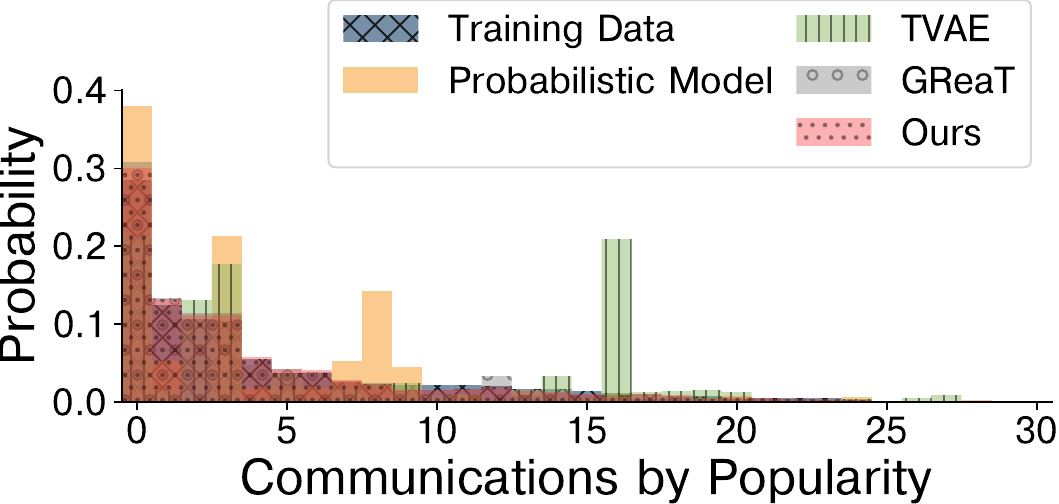}
        \caption{\small Distribution of popular edges.}
        \label{fig:dist_by_edges}
    \end{subfigure}
    \begin{subfigure}{.39\textwidth}
        \includegraphics[width=\linewidth]{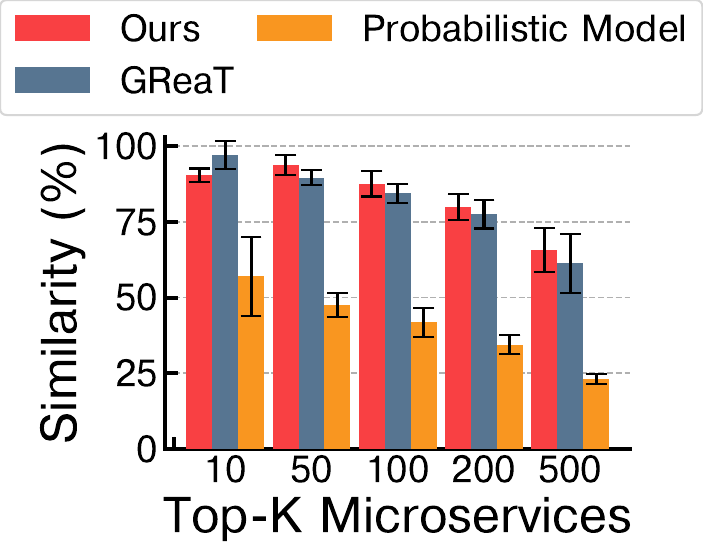}
        \caption{\small Heavy-hitter prediction.}
        \label{fig:heavy_hitter_acc}
     \end{subfigure}
    \caption{\small Distribution similarity between real and synthetic traces.}
    \label{fig:dist_similarity}
\end{minipage}
\hfill
\begin{minipage}[]{0.3\textwidth}
    \centering
    \includegraphics[width=\linewidth]{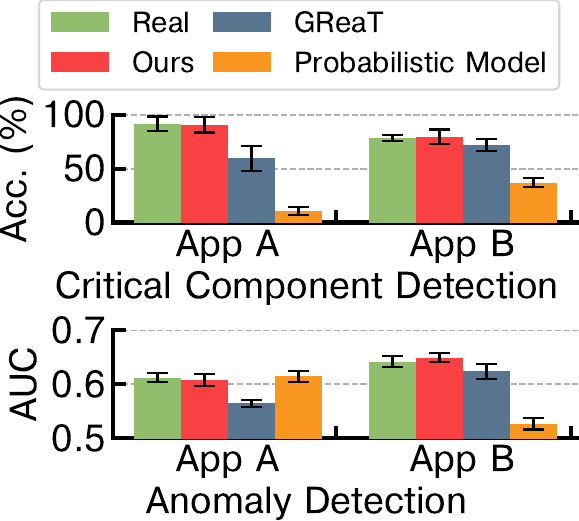}
    \caption{\small ML Model Performance (real vs. synthetic traces).}
    \label{fig:ml_performance}
\end{minipage}
\vspace{-0.6cm}
\end{figure*}

\mypara{Distribution of Popular Calls}
Realistic synthetic traces should mirror real-world communication patterns. To assess this, we analyze the distribution of calls, defined by \textit{(Source, Destination, Communication type)}. \autoref{fig:dist_by_edges} compares the 100 most popular calls generated by our method and the baselines, displaying the top 30 for clarity.

The KL divergence for traces generated by LLM-based approaches (ours and GReaT) is 0.16 and 0.11 respectively, indicating close similarity to the training data, whereas the probabilistic model's divergence is significantly higher at 3.84, due to its random selection processes. TVAE shows an intermediate divergence of 0.74, which is better than the probabilistic model but still less accurate than our method in capturing popular call distributions.

\mypara{Heavy-hitter Prediction}
Generating top-$K$ heavy-hitter microservices—those most frequently triggered in a sequence of call graphs—is crucial for resource optimization and anomaly detection in microservice management. In this experiment, we select 1K validation traces and create instructions with a service ID and call graph attributes like depth and \#edges to guide trace generation for both our model and the baseline. Similarity is evaluated by comparing the top-$K$ microservices between synthetic and validation traces over 20 runs.

\autoref{fig:heavy_hitter_acc} illustrates the similarity for varying $K$ values, from 10 to 500. Our method achieves over 90\% accuracy for $K\leq50$ and 65\% at $K$=500, demonstrating robust performance. 
The model trained with the GReaT method also shows robust performance, but slightly worse performance with larger $K$ values. We believe that the performance gap results from the lack of capability to generate complex structures (\autoref{subsec:structured_reasoning}), which impacts trace distribution.
In contrast, the probabilistic model starts at 59\% accuracy for $K$=10 and declines to 23\% at $K$=500, showcasing our method's capability to capture and replicate heavy-hitter dynamics.
% \aditya{if this is the only place we are evaluating GREAT then that's problematic as GREAT comes across as well performing enough}

Additional evaluation on branching (in/out-degree) and latency distributions is in~\autoref{appendix:more_dist_similarity}.

\subsection{Synthetic Data as Training data for ML-Driven Microservice Management}\label{subsec:use_cases}

Synthetic datasets can be used as a substitute for scarce real data in the training process for ML-driven microservice management tasks. Thus, we assess how well microservice management tasks using ML models for critical component extraction in FIRM~\citep{firm} and anomaly detection in TraceVAE~\citep{tracevae} perform when the models are trained on the synthetic datasets. The ML models are evaluated using real test data, and their results are compared to their original performance when trained on the real training dataset.

When choosing training data, we select a subset of traces from real data and label them with corresponding conditions (e.g., critical microservices). Then, we extract instructions from real data and use them to generate synthetic traces. We exclude invalid call graphs using the same accuracy metrics in~\autoref{subsec:structured_reasoning} before training. We train the models on 5K synthetic call graphs and test on 2K real call graphs, using the same test dataset across all experiments. For baselines, we use synthetic traces generated by GReaT and the Alibaba probabilistic model. Each experiment is run 5 times, varying random seeds, and results are averaged.

\mypara{Critical Component Extraction}
FIRM~\citep{firm} predicts critical components (microservices likely to violate service level objectives (SLO)) using support vector machines (SVMs) trained on latency-related features from call graphs.
For our evaluation, we train SVMs to detect critical components using two popular applications (apps A and B) from our trace dataset. For each application, we randomly sample call graphs and train two SVMs: one with real data and one with synthetic data generated by our fine-tuned model.
As shown in \cref{fig:ml_performance}, SVMs trained on synthetic data achieve near-identical accuracy to those trained on real data, with a difference of less than 1.5 percentage points.
In contrast, SVMs trained on synthetic traces from baselines show a performance gap of 6 to 81 percentage points.

\mypara{Anomaly Detection}
For operators to efficiently diagnose system failures, anomaly detection models predict whether microservice call graphs include anomalous characteristics like irregular graph structure or time.
We assess our synthetic data quality using TraceVAE~\citep{tracevae}, a variational autoencoder (VAE) model that detects anomalous microservices in terms of time consumption.
We train TraceVAE models using real and synthetic trace data, similar to our previous experiment. 
\cref{fig:ml_performance} shows that models trained on synthetic traces from our method yield results comparable to those trained on real data, as measured by ROC AUC.

We obtained similar results with two other classification tasks using fine-tuned Llama-2 7B models, as detailed in~\autoref{appendix:more_usecases}.

\subsection{Instruction-following Capability}
\label{subsec:instruction}

Enabling users to specify desired characteristics of synthetic data is crucial for trace generation. Such "custom" traces are useful to study corner cases and debug microservice management techniques. We assess our instruction-tuned model's capacity to accurately generate call graphs with specified attributes, such as high latency and rare communication types. We also explore the model's performance when prompted with combinations of these attributes that were not included in the training data.

When constructing the instruction tuning training datasets, we embed specific instructions to guide the generation of call graphs:

\begin{packeditemize}
    \item \textbf{High Latency:} Instructions specify that call graphs should exhibit latencies above the 90th percentile (p90) of the training dataset's distribution, varying by service. For example: \texttt{Build a call graph with high latency}.
    \item \textbf{Uncommon Communications:} Instructions indicate that the call graph layer should include a communication occurring in less than 10\% of the training data. For example: \texttt{Include an edge from (SRC) to (DEST) with (TYPE) communication type}.
\end{packeditemize}

We avoid combining these specific instructions in training samples to test the model's response to novel instruction combinations during inference.

\begin{figure}
    \begin{minipage}[b]{0.235\textwidth}
    \centering
    \includegraphics[width=\linewidth]{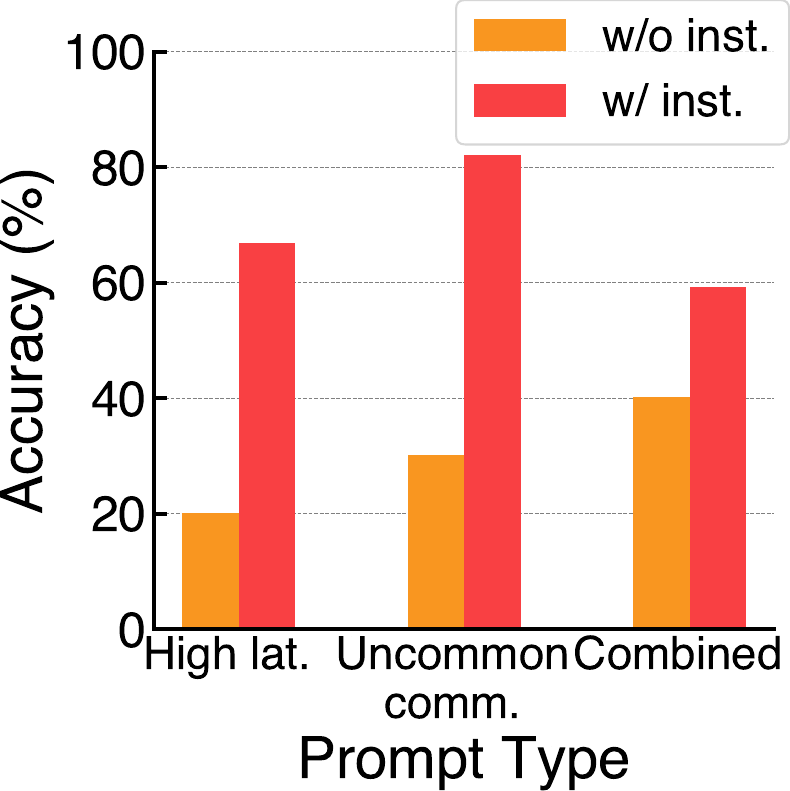}
    \caption{\small Instruction following accuracy (\%).}
    \label{fig:inst_following_accuracy}
    \end{minipage}
    \hfill
    \begin{minipage}[b]{0.235\textwidth}
    \centering
    \includegraphics[width=\linewidth]{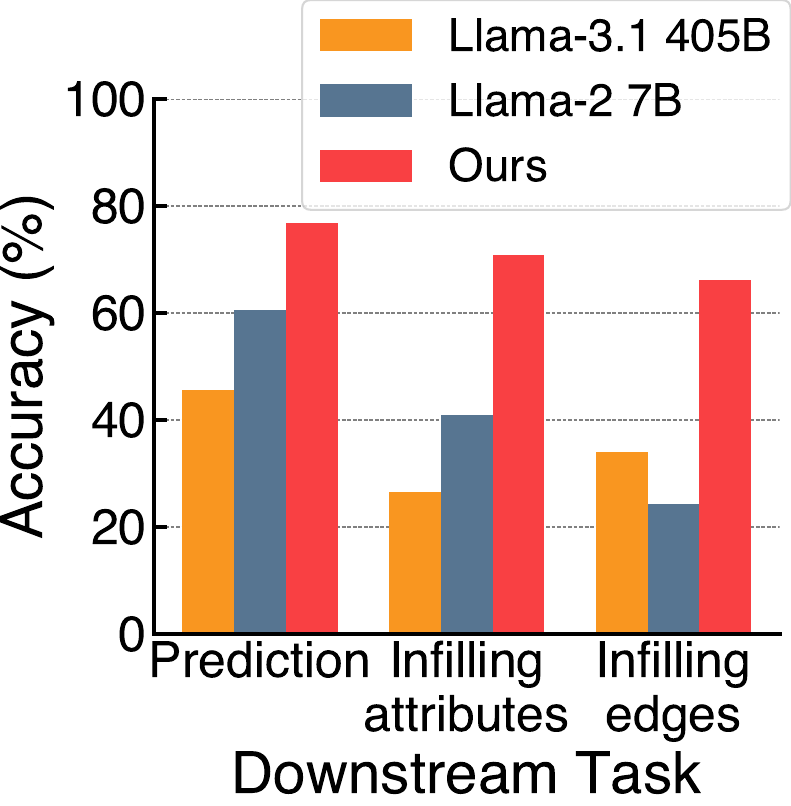}
    \caption{\small Downstream task accuracy (\%).}
    \label{fig:dowmstream_task}
    \end{minipage}
\vspace{-0.6cm}
\end{figure}

\mypara{Results}
\autoref{fig:inst_following_accuracy} presents the instruction-following accuracy for high latency and uncommon communication. We assessed this by filtering 1K validation instructions to see how many generated call graphs met the defined criteria (\eg exceeding p90 latency). We also compared these results against outputs generated without specific instructions to assess the impact of tailored prompts.

Additionally, we evaluate the model's performance when both instructions were combined, a scenario not covered during training. The model's ability to satisfy both conditions simultaneously, despite not being explicitly trained to do so, is detailed in the right of~\autoref{fig:inst_following_accuracy}. Higher accuracy in the absence of instructions may arise from inherent biases, such as those related to service ID or edge count, which align with the desired outcomes.

\subsection{Adapting Models for Trace-Related Downstream Tasks}\label{subsec:downstream_tasks}
We extend our evaluation beyond generating synthetic traces, demonstrating the utility of our pre-trained model in performing downstream tasks related to microservice traces. The trace pre-trained model is adapted to each downstream task through additional fine-tuning.
We focus on scenarios where partial information from distributed environment traces is available, emphasizing the challenges posed by incomplete data. This section compares our fine-tuned model with the standard Llama-2 7B, which lacks specific training on call graph data, and with Llama-3.1 405B by providing task descriptions and up to 16 examples in prompts (\ie in-context learning~\citep{gpt3}), to highlight the need for domain-specific training.

\mypara{Predicting Uncommon Communications}
The task is to predict uncommon communication patterns (as in~\autoref{subsec:instruction}) based on the first 10 lines of a trace.
For fine-tuning, we adapt both the original Llama-2 7B and our trace pretrained model
to this binary classification task using 15K samples. 
Each sample's prompt comprised the first 10 edges of a real trace, with binary labels indicating the presence of uncommon communication patterns in the subsequent trace sections.

As shown on the left side of \autoref{fig:dowmstream_task}, the original Llama-2 model achieves only 60.6\% accuracy, indicating insufficient training for recognizing uncommon patterns.
Additionally, in-context learning with Llama-3.1 405B shows lower accuracy (45.6\%), suggesting 
that larger models trained on general internet data struggle with domain-specific tasks.
In contrast, our model achieves 76.8\% accuracy, demonstrating its enhanced capability to interpret and predict based on partial trace data.

\mypara{Infilling Missing Data}
Missing data is common in large-scale trace logging, such as in Alibaba's microservice call graphs, where 67\% of traces contain missing values~\citep{alibaba_inconsistency}. This task focuses on fine-tuning our model to accurately infill missing data in microservice call graphs, considering partial information.
Specifically, we conduct two experiments on infilling (1) a missing attribute and (2) a missing call connecting two layers.

In the first experiment, we construct a training dataset with 1.2K questions, each containing a sequence of edges with one attribute marked as \texttt{[MISSING]}. The missing value is the unknown ground truth for prediction, so these are multi-class classification problems.
Attributes targeted include communication type (\eg HTTP, RPC) or destination microservice. 
We evaluate the model on a 6K-sample test dataset, where our model demonstrated over 70\% accuracy in predicting the correct attributes, significantly outperforming the accuracy of baselines by about 30\% to 40\% as reported in the middle of \autoref{fig:dowmstream_task}.

The second experiment's dataset comprises 1K samples, each representing a pair of parent and child layers with a missing connecting edge tagged as \texttt{[MISSING]}. 
After training, we evaluate both models on 5K test cases to generate the correct edge, ensuring the finish time matched or exceeded the start time. 
The right part of \autoref{fig:dowmstream_task} shows that while the original Llama-2 model scored only 24\% accuracy and Llama-3.1 405B reached 34\%, our model maintained a high accuracy of 66\%, underscoring its robustness in more complex tasks.

These experiments demonstrate the capabilities of our trace pre-trained model to effectively adapt to handle infilling tasks that even large foundation models like Llama-3.1 405B cannot achieve.

%% file: conclusion.tex
\section{Conclusion} \label{sec:conclusion}
This paper introduces \sys{}, a training method for adapting LLMs to generate microservice trace graphs using recursive call graph generation and instruction tuning. Our approach outperforms baselines in producing accurate, valid call graphs with improved distributional similarity to real traces. We demonstrate that synthetic traces can effectively replace real data for training microservice management tasks, such as critical component extraction and anomaly detection. Additionally, instruction tuning enhances graph generation based on user-specified features, enabling applications in prediction and data infilling. While we focus on microservice call graphs, our method broadly applies to other system traces with similar structures.  

\mypara{Limitations}
The recursive method improves the accuracy of call graph generation compared to generating the entire trace at once, but a key drawback is that previously generated edges are discarded, as only the conditioning information from the prior layer is passed to the next layer generation steps.
Although dropping previously generated edges has little impact on the output in microservice call graph generation, where direct neighbors exert the most influence~\citep{ursa}, incorporating past information, such as prior layers or a time series of call graph traces, could enhance the capture of longer-range dependencies and temporal patterns.
However, efficiently compressing historical trace information while preserving critical details remains an open challenge. In future work, we will consider this direction to compress long-range traces and generate synthetic traces conditioned on the compressed traces.

Furthermore, our method uses manually constructed instruction templates, which may lead to suboptimal generation quality, as we are not using the full potential of language models pre-trained with trillions of tokens~\citep{touvron2023llama}.
Following the methods of prior work~\citep{llava, phi, li2024selective}, we believe that diversifying instructions using LLM-generated output is a potential method to improve the ability of LLMs to follow user intentions.
However, naively guiding LLMs to generate instructions for trace generation may result in instructions that lack useful characteristics for downstream tasks. In future work, we plan to integrate domain-specific knowledge of traces to improve the usefulness and diversity of instructions generated by LLMs.

Lastly, we focused on generating microservice call graphs in this paper, but other system traces, such as operating system~(OS) call graphs, share a similar hierarchical structure. The primary differences in OS call graphs lie in their greater depth and the increased diversity of node and edge types.
We present relevant experimental results in~\autoref{appendix:batch_job_gen}, using cluster batch job traces.
We believe that extending \sys’s capabilities to diverse types of system traces is important, since \sys{} can function not only as a synthetic trace generator but also as a world model~\citep{kimik2} that offers realistic feedback for developing more sophisticated agents capable of reasoning over various computer system traces.

\mypara{Ethics Statement} There are no ethical concerns raised by our work as the data used in this study is public with sensitive information redacted.

\mypara{Acknowledgments} This material is based upon work supported by the U.S. National Science Foundation (NSF) under Grant Number 2326576. We acknowledge the use of AI assistants to enhance writing clarity.

%% file: appendix.tex
%%%%%%%%%%%%%%%%%%%%%%%%%%%%%%%%%%%%%%%%%%%%%%%%%%%%%%%%%%%%%%%%%%%%%%%%%%%%%%%
%%%%%%%%%%%%%%%%%%%%%%%%%%%%%%%%%%%%%%%%%%%%%%%%%%%%%%%%%%%%%%%%%%%%%%%%%%%%%%%
% APPENDIX
%%%%%%%%%%%%%%%%%%%%%%%%%%%%%%%%%%%%%%%%%%%%%%%%%%%%%%%%%%%%%%%%%%%%%%%%%%%%%%%
%%%%%%%%%%%%%%%%%%%%%%%%%%%%%%%%%%%%%%%%%%%%%%%%%%%%%%%%%%%%%%%%%%%%%%%%%%%%%%%

\newpage
\appendix

\input{relatedwork}

\section{Training Details}\label{appendix:training_details}

\subsection{Training Setup}

We train all models with 4$\times$ A100 80GB GPUs in our cluster with the hyperparameters described in~\autoref{tab:hyperparameters}.
We apply LoRA~\citep{hu2022lora} adapters to query and key projection matrices of attention layers with $rank=8$, $alpha=16$, and $dropout=0.1$.
For the downstream task training in~\autoref{subsec:downstream_tasks}, we freeze the backbone model and only train the last classification layer for the prediction task. For the infilling downstream task, we use LoRA adapters with the same configuration as mentioned earlier.
During inference, we use a temperature of 0.8 and top-K of 50 for trace generation, unless otherwise specified.
We use models (Llama 2 and 3.1) under appropriate community license.

\begin{figure}[]
  \centering
  \begin{subfigure}{.49\textwidth}
   \includegraphics[width=\linewidth]{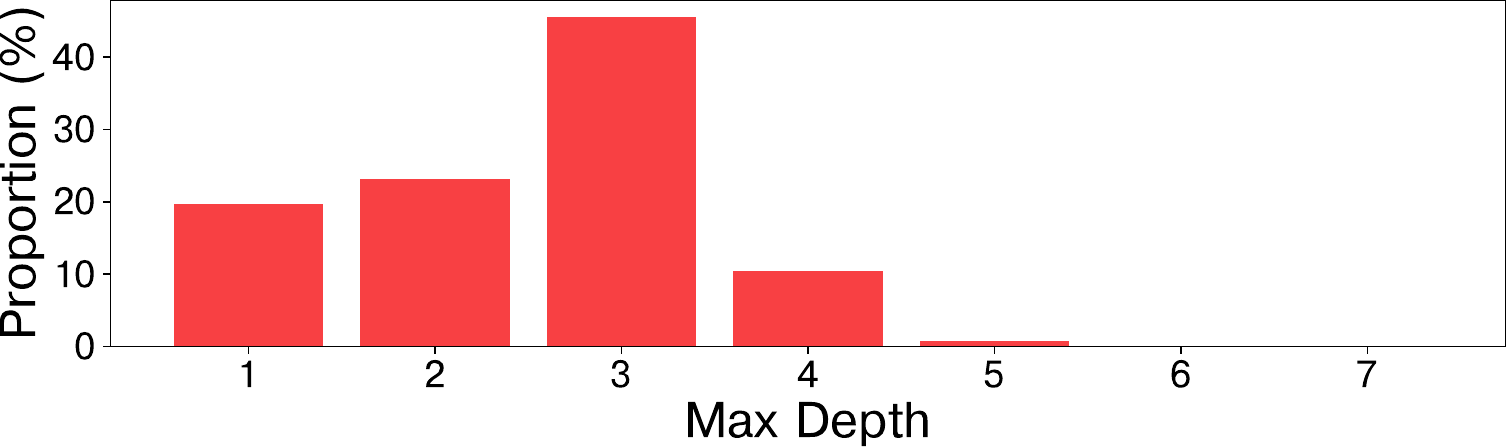}
    \caption{Distribution by call graph depth.}
    \label{fig:training_cg_depth_dist}
  \end{subfigure}
  \begin{subfigure}{.49\textwidth}
   \includegraphics[width=\linewidth]{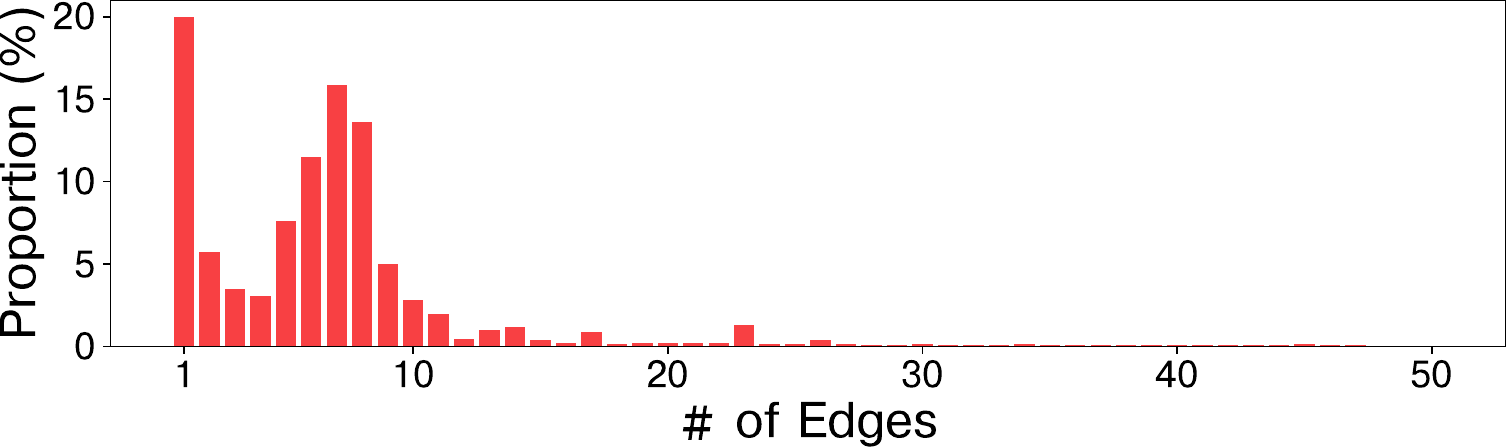}
    \caption{Distribution by the number of edges.}
    \label{fig:edge_dist}
  \end{subfigure}
  \caption{Training data distribution after preprocessing.}
  \label{fig:training_data_dist}
\end{figure}

\begin{table*}[]
\small
\centering

\begin{tabular}{@{}cll@{}}
\toprule
\textbf{Model}                  & \textbf{Hyperparameter} & \textbf{Value} \\ \midrule
\multirow{8}{*}{Pre-Training \& Instruction Tuning} & Optimizer & AdamW~\citep{adamw} \\ \cmidrule(l){2-3} 
                      & Learning rate & 3e-4 with cosine scheduler \\ \cmidrule(l){2-3} 
                      & Batch size &  64 \\ \cmidrule(l){2-3} 
                      & Gradient clipping &  1.0 \\ \midrule
\multirow{8}{*}{Downstream Task Fine-tuning} & Optimizer  & AdamW \\ \cmidrule(l){2-3} 
                      & Learning rate & 1e-4 with cosine scheduler \\ \cmidrule(l){2-3} 
                      & Batch size & 2 \\ \cmidrule(l){2-3} 
                      & Gradient clipping & 1.0 \\ \bottomrule
\end{tabular}
\caption{Training setup and hyperparameters.}
   \label{tab:hyperparameters}
\end{table*}

\begin{figure*}
  \centering
  \includegraphics[width=\linewidth]{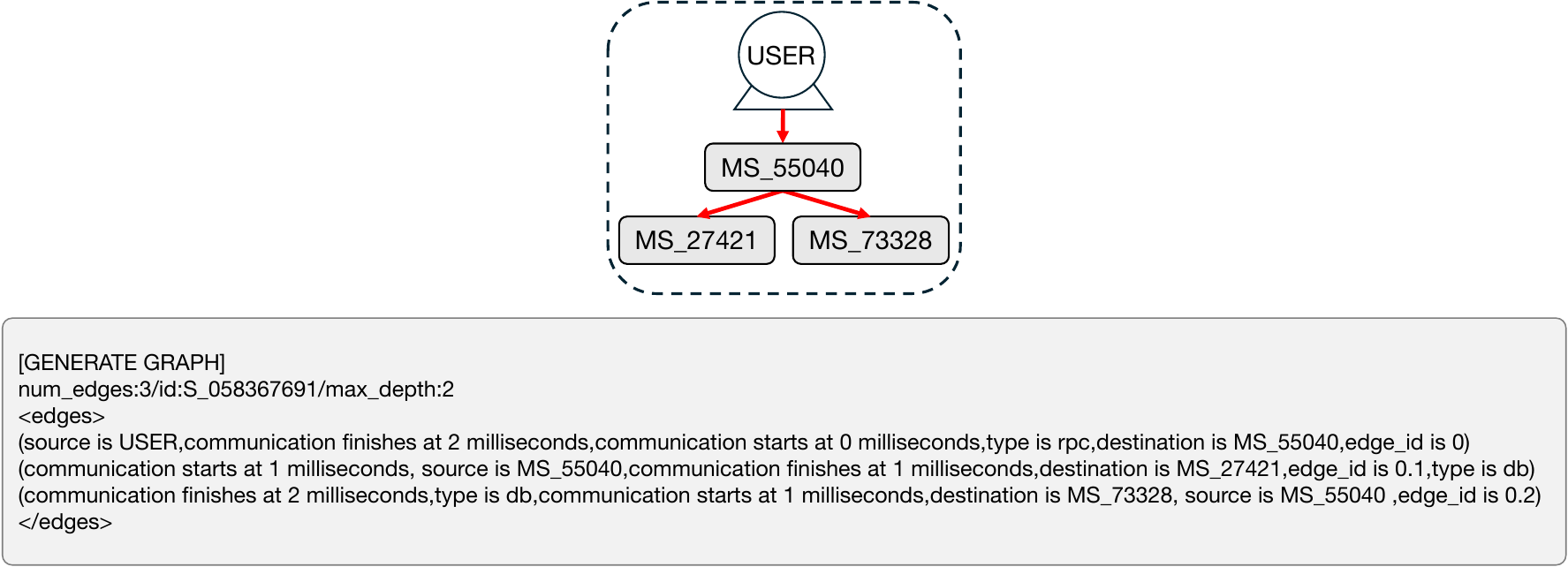}
   \caption{A training data sample of a call graph with 3 edges represented in tabular format.}
   \label{fig:baseline_training_example}
 \end{figure*}

  \begin{figure*}
    \centering
    \includegraphics[width=\linewidth]{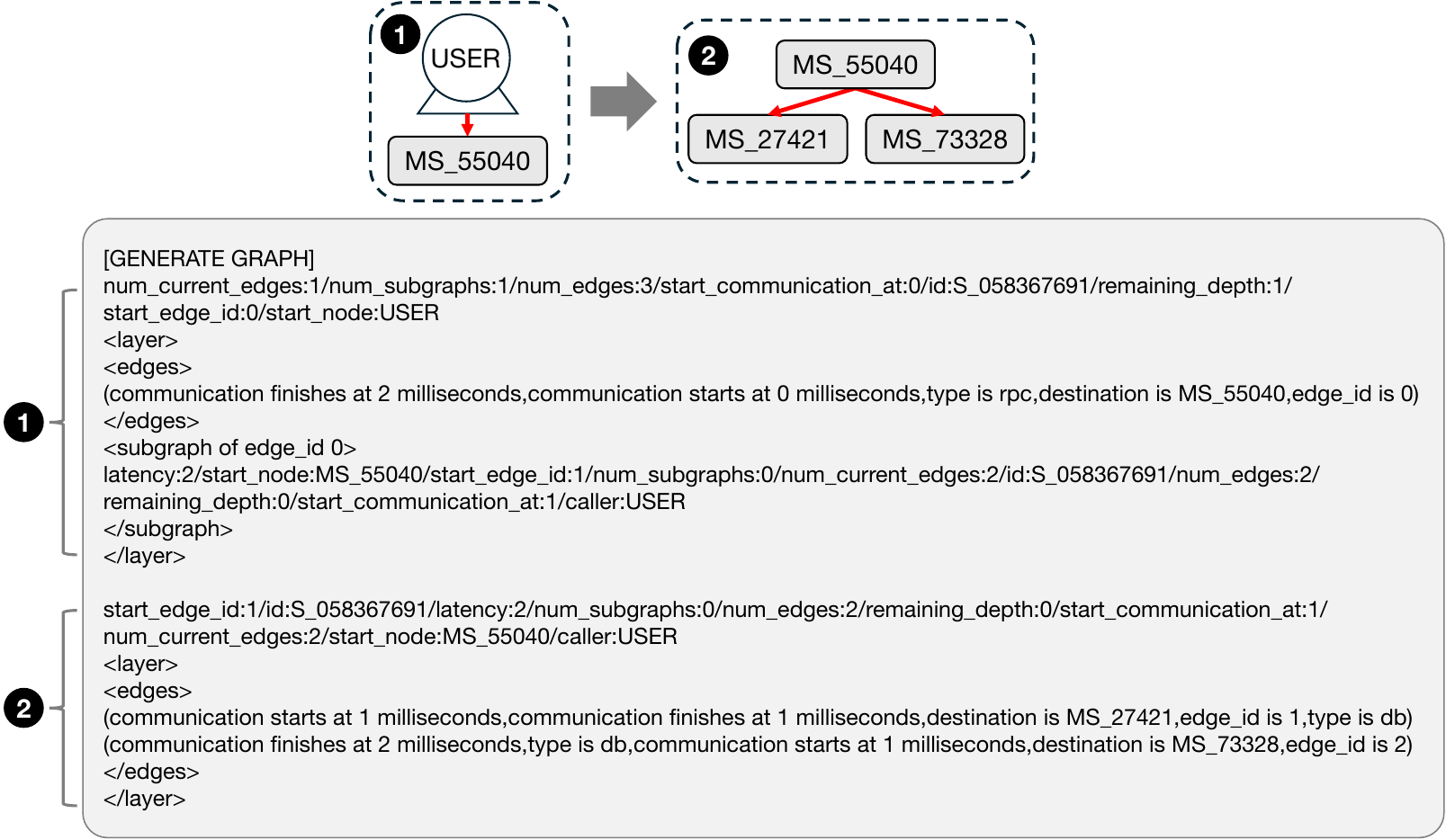}
     \caption{A training data sample of a call graph with 3 edges for recursive generation.}
     \label{fig:recursive_training_example}
   \end{figure*}

  \begin{figure*}
    \centering
    \includegraphics[width=\linewidth]{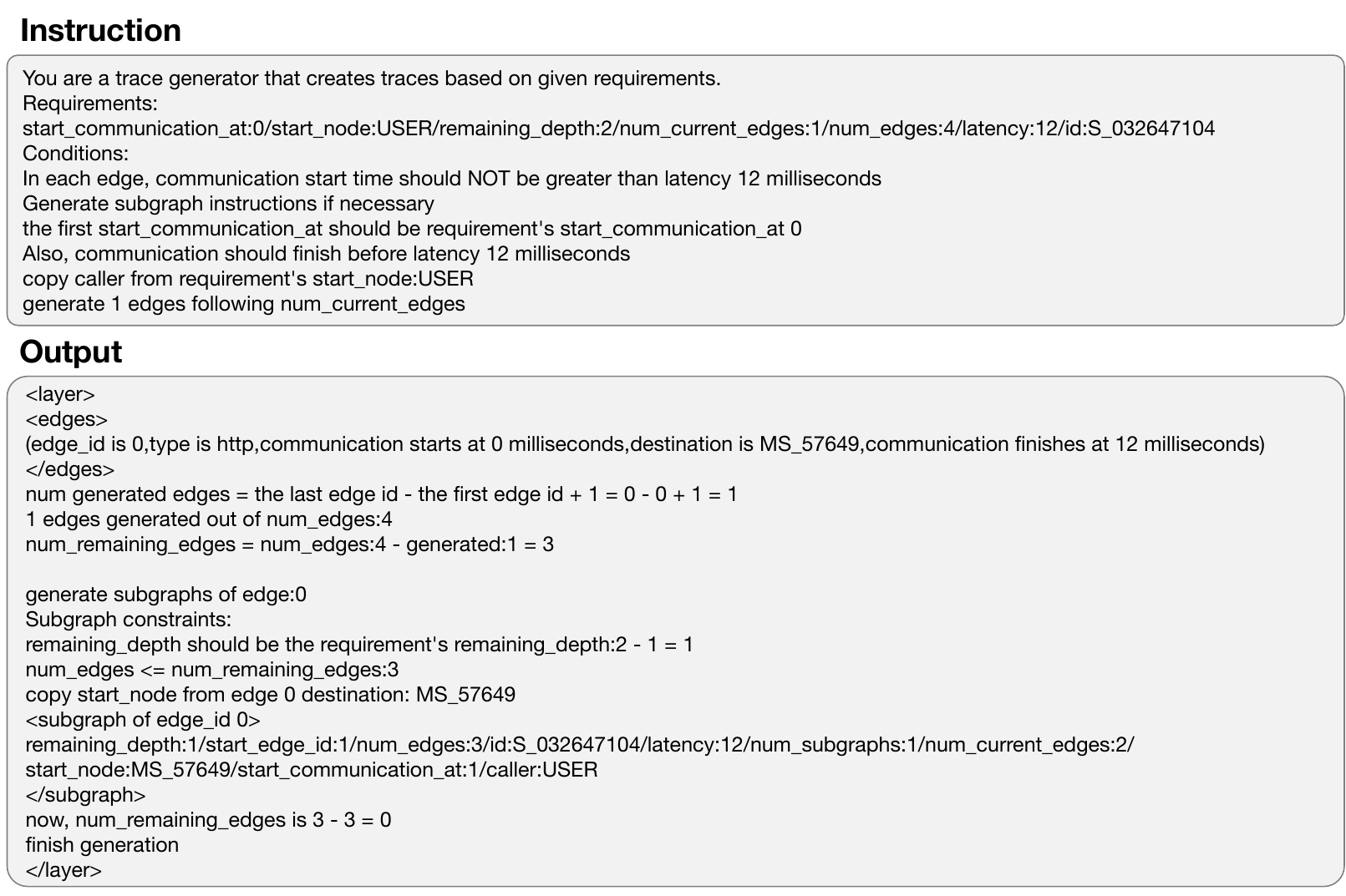}
     \caption{A training data sample of a call graph layer for instruction-tuning.}
     \label{fig:inst_tuning_example}
   \end{figure*}
   
\subsection{Training Data Preprocessing}\label{appendix:training_data_preprocessing}
From the Alibaba microservice v2022 traces~\citep{madu}, we use the first-hour call graph traces as our training data, which consist of 6434 unique microservices collected from more than 10 clusters.
We reserve 10\% of these samples for validation.
The traces are collections of API calls, where each API call includes communication information between the two microservices.
Service ID is a nine-digit number starting with the prefix "S\_" instead of using a real service name (e.g., social network), and microservice is a five-digit number starting with the prefix "MS\_".
We construct call graphs using the \texttt{trace ID} field (i.e., API calls with the same trace ID belong to one call graph).
When constructing call graphs, we remove calls with missing information (e.g., destination microservice IDs are unknown) and remove call graphs that are not connected (e.g., missing edges).
To remove redundancy, we filter out call graphs that have the same structure and fields (e.g., service ID, latency) for all API calls.
The distributions of training data after removing redundancy are shown in~\autoref{fig:training_data_dist}.

Note that the dataset anonymizes service and microservice names, ensuring our model does not disclose sensitive information.
However, if the training dataset contains sensitive data, models trained with our method may still risk exposing privacy-related information.

The instruction-tuning datasets were created by randomly selecting 5\% of the training graphs and reformatting them for instruction tuning. The instruction-tuning lasted four epochs.

\subsection{Training Data Examples}\label{appendix:training_data_example}
From the call graph traces, we create text-based representations of call graphs as described in~\cref{subsec:encoding_call_graphs}.
First of all, \autoref{fig:baseline_training_example} is a training data example of converting a call graph into a tabular data format, which is the baseline in~\autoref{subsec:structured_reasoning}.
At the beginning, we provide high-level information about the call graph, including the service ID, number of edges, and graph depth.
Each line inside the \texttt{<edges>} block corresponds to a call in a call graph.
6 fields exist for each call including the edge ID, source/destination microservices, communication type, and communication start/finish time.

\autoref{fig:recursive_training_example} shows an example training data sample for recursive generation as described in~\cref{sec:recursive}.
Each sample consists of a sequence of layers, where each layer includes the edges and the conditions for the next layers.
At the beginning of each layer, we provide high-level information to explain connections with the previous layers (e.g., \texttt{start\_node}, \texttt{caller}), structure in the call graph (e.g., \texttt{remaining\_depth}, \texttt{num\_edges}, \texttt{start\_edge\_id}), and time-related information (e.g., \texttt{latency}, \texttt{start\_communication\_at}).
Note that the number of fields in each edge is reduced from 6 to 5 since the edges share the same start node.
Also, the edge ID field is an integer, not a dot-decimal number.
For each next layer, the condition is described in each \texttt{<subgraph>} block starting with the edge ID to be extended.

\autoref{fig:inst_tuning_example} is an example of instruction-tuning data.
The instruction starts with a system prompt followed by conditions as in~\autoref{fig:recursive_training_example}.
We further explain the condition in natural language along with user-requested features, as studied in~\autoref{subsec:instruction}.
In the output section, we include Chain-of-Thought scratchpads at the end of \texttt{<edges>} block and at the beginning of \texttt{<subgraph>} blocks, which elaborate on the number of edges to generate and constraints of subgraph conditions.
For example, the scratchpad includes descriptions of the depth requirement to help LLMs understand that the depth field should be reduced by 1 from the current layer's depth.

As described in~\cref{subsec:encoding_call_graphs}, we drop each call graph attribute randomly with probability $p_{drop}$.
We set $p_{drop}$ to 0.9 for all attributes except for the service ID field, which is always kept ($p_{drop}=1$), to ensure minimal conditioning.

\begin{figure}[]
    \centering
    \begin{subfigure}{\linewidth}
        \centering
        \includegraphics[width=\linewidth]{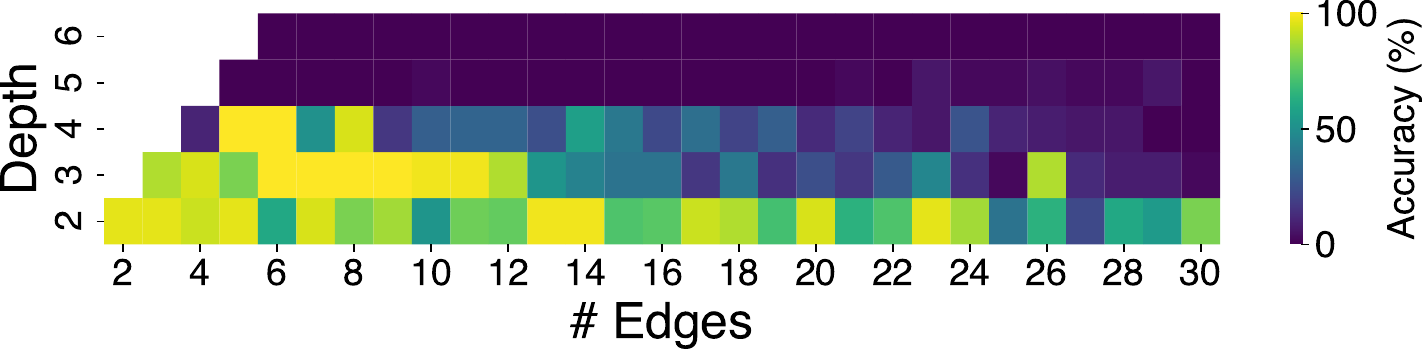}
        \caption{Baseline accuracy heatmap.}
        \label{fig:heatmap_baseline}
    \end{subfigure}
    \begin{subfigure}{\linewidth}
        \centering
        \includegraphics[width=\linewidth]{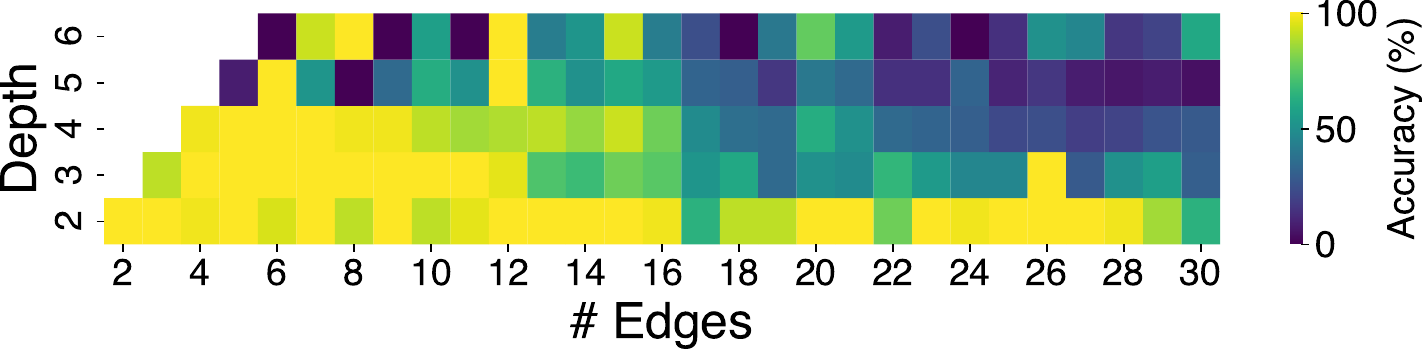}
        \caption{Accuracy heatmap with recursive generation.}
        \label{fig:heatmap_recursive}
    \end{subfigure}
    \begin{subfigure}{\linewidth}
        \centering
        \includegraphics[width=\linewidth]{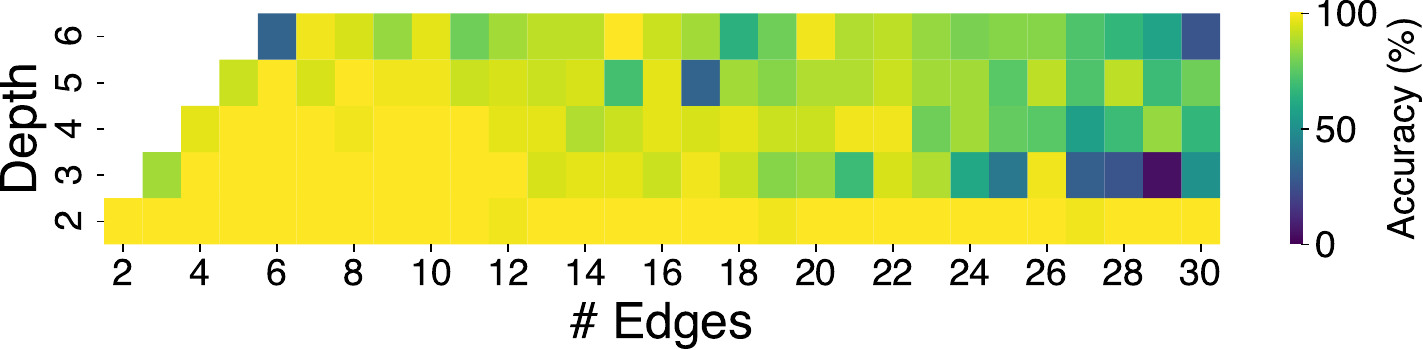}
        \caption{Accuracy heatmap with recursive generation and instruction-tuning.}
        \label{fig:heatmap_recursive_inst}
    \end{subfigure}
    \caption{Accuracy heatmap.}
    \label{fig:heatmap}
\end{figure}  

\section{Constraints in Call Graph Layers}\label{appendix:validation_rules}
In this section, we describe constraints to be met for each generated call graph layer to be correct.
First of all, the generation results are considered invalid if the output does not have the valid format with \texttt{<edges>} and \texttt{<subgraph>} tags.

\mypara{Edges}
For each edge, we check the following conditions.
First of all, each edge should include the 5 fields: edge ID, destination, communication type, and communication start/finish time.
Secondly, we check whether the right number of edges are generated as described in the condition.
Third, the communication start time should be equal to or greater than the communication start time described in the condition, and should not be greater than the communication finish time of the edge.
Lastly, the communication finish time must not exceed the latency specified in the condition.

\mypara{Next Layer Conditions}
For the next layer conditions, we first check whether the next layer conditions should be generated or not.
If the remaining depth field in the instruction is 0 or the number of edges that need to be generated is 0, no \texttt{<subgraph>} blocks should be generated.

Then, we check the validity of each field in the next layer conditions.
First of all, the edge ID inside the \texttt{<subgraph>} block should be found in the edges generated in the current layer.
For the depth, the remaining depth field should be less than the remaining depth of the instruction.
Additionally, at least one of the resulting subgraphs must have a depth that is reduced by one compared to the original graph.
For the \texttt{start node} and \texttt{caller} fields, they should be copied from the destination from the parent edge and the start node from the instructions, respectively.
Lastly, we check the latency and communication start time by comparing the values to those of the parent edge.
The latency of a child layer should not be greater than the communication finish time of the parent edge.
Also, the communication start time of a child layer must not be earlier than that of the parent edge.

After generating both edges and the next conditions, we check if the sum of the number of edges matches the number of edges in the instruction.

\section{Additional Evaluation Results}\label{appendix:eval}

\subsection{Structured Reasoning Results in Detail}\label{appendix:accuracy_detail}

\begin{figure}[]
    \centering
    \includegraphics[width=0.8\linewidth]{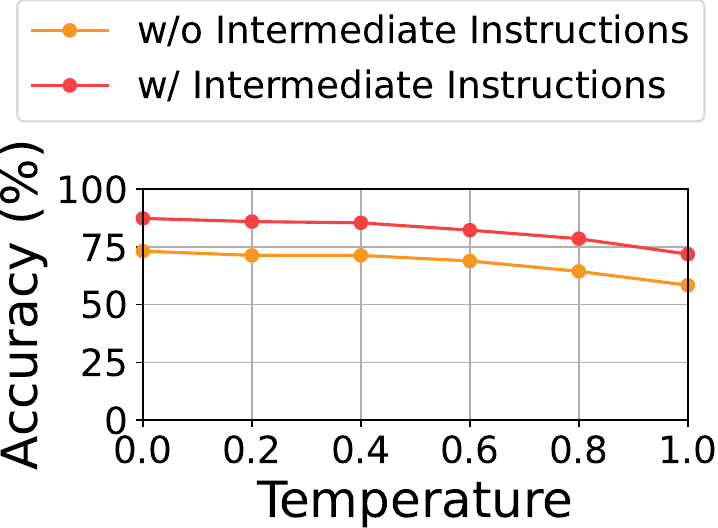}
     \caption{Accuracy to generate correct call graph structures with and without intermediate instructions during instruction tuning.}
     \label{fig:accuracy_with_cot}
\end{figure}

\begin{figure*}
    \centering
    \begin{subfigure}{.45\textwidth}
        \includegraphics[width=\linewidth]{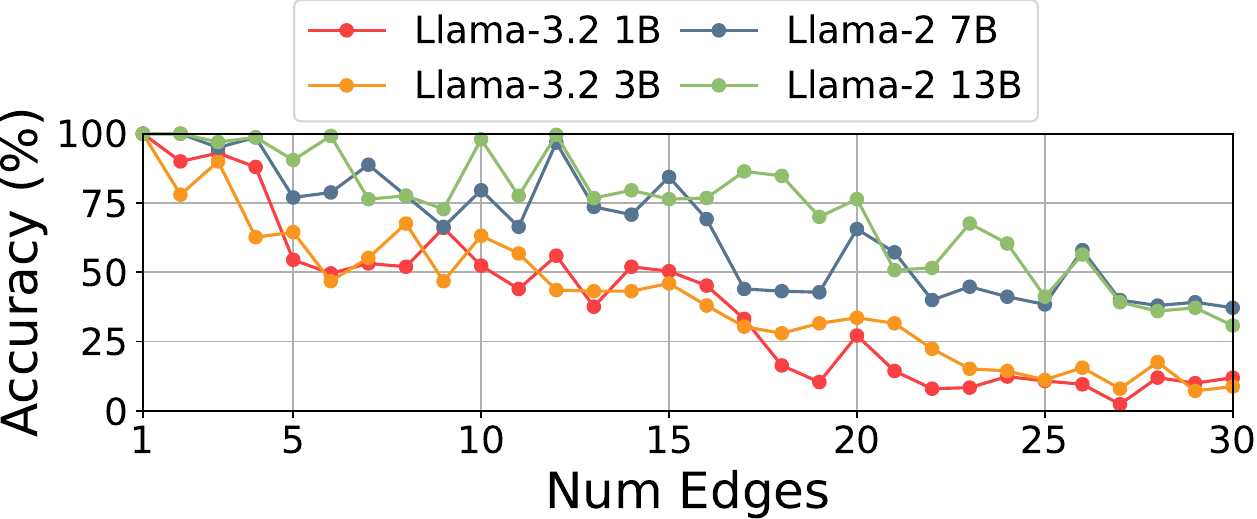}
        \caption{accuracy vs. number of edges}
        \label{fig:accuracy_by_edges_model_size}
    \end{subfigure}
    \begin{subfigure}{.25\textwidth}
        \includegraphics[width=\linewidth]{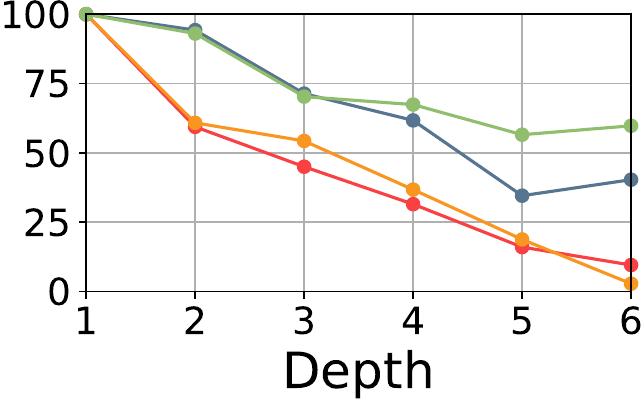}
        \caption{accuracy vs. depth}
        \label{fig:accuracy_by_depth_model_size}
    \end{subfigure}
    \begin{subfigure}{.25\textwidth}
        \includegraphics[width=\linewidth]{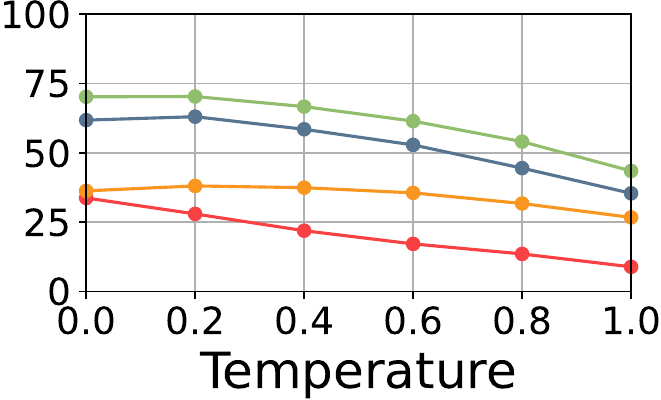}
        \caption{accuracy vs. temperature}
        \label{fig:accuracy_by_temp_model_size}
     \end{subfigure}
    \caption{\small Call graph generation accuracy with varying model sizes. The plots in (a) and (b) show the accuracy varying edges and depth using greedy sampling, and (c) shows the accuracy varying sampling temperature.}
    \label{fig:accuracy_varying_model_size}
\end{figure*}

This section provides a more detailed analysis of the results from~\autoref{subsec:structured_reasoning}, accuracy to generate call graphs adhering to all structural constraints while matching the specified attributes in prompts (i.e., \texttt{num\_edges} and \texttt{depth}).
\autoref{fig:heatmap} offers a closer look at~\autoref{fig:accuracy_by_edges} and~\autoref{fig:accuracy_by_depth}, where each grid point $(X, Y)$ represents accuracy for prompts with $X$ edges and a maximum depth of $Y$.
\autoref{fig:heatmap_baseline}, \autoref{fig:heatmap_recursive}, and \autoref{fig:heatmap_recursive_inst} correspond to the same settings as \textit{(baseline)}, \textit{(recursive)}, and \textit{(recursive + instruction)} from~\autoref{subsec:structured_reasoning}, respectively.
The results in \autoref{fig:heatmap} show that the recursive generation and instruction tuning improves accuracy across most combinations of \texttt{(\# Edges, Depth)}.
However, some configurations in~\autoref{fig:heatmap_recursive} and~\autoref{fig:heatmap_recursive_inst} exhibit lower accuracy, likely due to the distribution of training data in terms of edge count and depth.

In addition, we conduct an ablation study, where we remove \textit{intermediate instructions} during instruction tuning to see the impact of \textit{intermediate instructions} in generating correct call graphs.
For instance, we remove equations and sentences that help to reason the properties to be generated (e.g., a sentence "\texttt{num generated edges = the last edge id - the first edge id + 1}" in \autoref{fig:inst_tuning_example}).
\autoref{fig:accuracy_with_cot} reports the call graph generation accuracy varying the sampling temperature.
Notably, removing the intermediate instructions during instruction tuning results in an approximate 13\% decrease in accuracy across all temperatures, demonstrating the effectiveness of having intermediate reasoning steps during instruction tuning.

\color{black}
\subsection{Structured Reasoning Results varying Model Sizes}\label{appendix:accuracy_varying model size}

To evaluate the impact of model size on trace generation performance, we report the generation accuracy of models with varying numbers of parameters. Specifically, we compare four models: Llama-3.2 1B, Llama-3.2 3B, Llama-2 7B, and Llama-2 13B. Each model undergoes pre-training~(\autoref{subsec:pre-training}) using the same training dataset~(same as the \textit{Recursive} setup described in \autoref{subsec:structured_reasoning}).

\autoref{fig:accuracy_varying_model_size} presents the microservice call graph generation accuracy across different model sizes. Overall, models with a larger number of parameters demonstrate higher accuracy, with this trend being particularly evident in~\autoref{fig:accuracy_by_temp_model_size}. Notably, models with more parameters perform better as the depth of prompts increases.
For instance, the 13B model achieves a 20 percentage point improvement over the 7B model for inputs with a depth greater than 4 as shown in~\autoref{fig:accuracy_by_depth_model_size}.

\subsection{Memorization}
\label{appendix:memorization}

\begin{figure}
  \centering
    \includegraphics[width=0.9\linewidth]{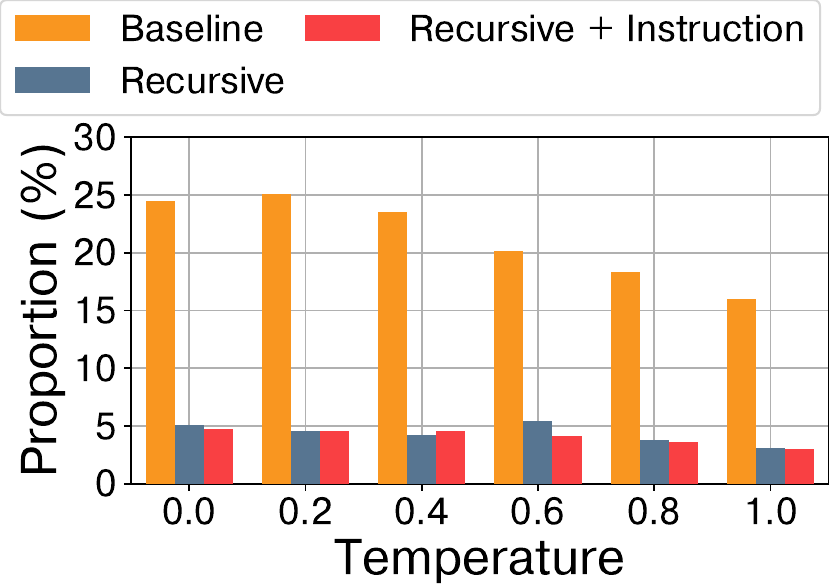}
    \caption{\small Proportion (\%) of synthetic call graphs found in training data varying temperature parameters.}
\label{fig:memorization}
 \end{figure}

We assess whether synthetic traces are generated by memorizing training data by measuring the percentage of traces that exactly match the structures and call graph attributes found in the training data. Specifically, for the synthetic traces generated in~\autoref{subsec:structured_reasoning}, we compute the proportion that exhibit identical call graph structures and edge attributes as those in the training data.

\autoref{fig:memorization} presents the proportion of memorized synthetic traces. Notably, traces generated using the \texttt{Baseline} method exhibit a relatively high level of memorization, with proportions ranging from 16\% to 24\%. In contrast, our methods (\texttt{Recursive} and \texttt{Recursive+Instruction}) demonstrate significantly lower memorization, with proportions ranging from 3\% to 5\%.

These results suggest that our recursive generation method is effective not only in producing more structured traces, as shown in~\autoref{subsec:structured_reasoning}, but also in minimizing the memorization of training data. This helps generate more diverse synthetic traces.

\color{black}

\subsection{More Experiments on Similarity Between Real and Synthetic Traces}
\label{appendix:more_dist_similarity}

To further evaluate the effectiveness of our method in capturing the complexity of microservice interactions, we analyze the distribution similarities of microservice branching and response times using 10K synthetic traces. For consistency, we include only correct call graphs in the evaluation, following the accuracy criteria outlined in~\autoref{subsec:structured_reasoning}.
The same baselines as in~\autoref{subsec:distribution_similarity} are used, including GReaT and Alibaba probabilistic model. To extend the probabilistic model to include time-related fields, we augment it to generate response times by sampling from the training data statistics.

\begin{figure}
  \centering
    \includegraphics[width=\linewidth]{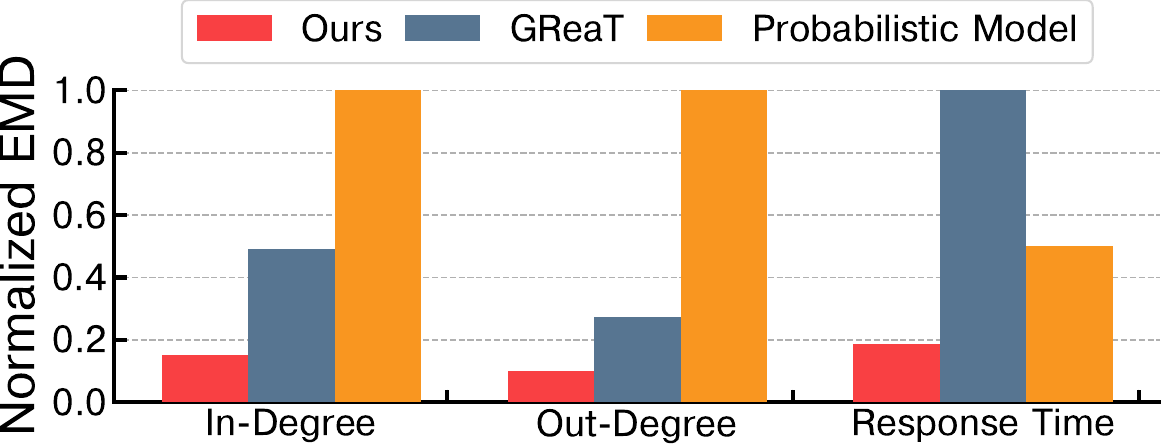}
    \caption{\small Distribution similarities in microservice branching (in-degree and out-degree) and response times between real and synthetic traces.}
\label{fig:more_dist_similarity}
 \end{figure}

\autoref{fig:more_dist_similarity} presents the distribution similarities for microservice branching and response times.
We use normalized Earth Mover's Distance (EMD) as the similarity metric, ensuring comparability across fields with varying scales.
\textit{In-Degree} represents the distribution of the number of communications received by each microservice, while \textit{Out-Degree} reflects the number of communications initiated by each microservice.
\textit{Response Time} measures the distribution similarity of the duration required to complete each communication.
Across all three metrics, our method consistently achieves the closest results to the training data, achieving a 2.6x to 10x reduction in EMD compared to GReaT and the probabilistic model.
We attribute its higher EMD values to an inability to generate complex call graph structures effectively.

\subsection{More Experiments on Using Synthetic Traces in ML Use Cases}\label{appendix:more_usecases}

\begin{table}[]
\centering
\small
\begin{tabular}{c|cc}
Accuracy (\%) & High Latency & \makecell{Uncommon \\ Communications} \\
\midrule
\makecell{Real}     & 68.3 \%      & 65.3 \%      \\
\makecell{Synthetic} & 67.1 \%      & 62.5 \%
\end{tabular}
\caption{Accuracy of prediction tasks by fine-tuning Llama-2 7B with real and synthetic traces.}
\label{tab:usecase_accuracy}
\end{table}

Building on the two evaluation tasks in \autoref{subsec:use_cases}, we conducted similar experiments using two classification tasks, fine-tuning the original Llama-2 7B models.
We predict high latency in call graphs, defined as latency equal to or above the 90th percentile for each service, without providing latency-related information in the input data. Secondly, we predict uncommon communications in direct neighbors within a call graph, as defined in~\autoref{subsec:instruction}.

We fine-tune the original Llama-2 7B as a classifier by replacing the last layer with a classification layer and training only the last layer for one epoch.
As in the experiments in~\autoref{subsec:use_cases}, we train one model using real and one using synthetic data. \autoref{tab:usecase_accuracy} reports the test accuracy on real test data. Although synthetic traces have a slight accuracy drop compared to real traces, they still exhibit similar characteristics and can be effectively used in real-world tasks.
For Llama-2 7B fine-tuning, We use a few thousand call graphs as training, validation, and test data (ratio 8:1:1) for each classification task and conduct a grid search over learning rates and batch size.

\color{black}

\subsection{Trace generation with reasoning models}

Recent reasoning models have demonstrated strong capabilities in solving complex tasks through structured thinking and reasoning. However, our preliminary experiments with reasoning models using few-shot settings did not improve trace generation accuracy.

\autoref{tab:reasoning_accuracy} shows the structured reasoning accuracy (Section 4.1) using a QwQ-32B~\citep{qwq32b} model, where we provide descriptions on call graph structures and constraints along with two randomly chosen trace examples in prompts. We generate 10 samples for each (num\_edges, depth) pair across ranges of 2 $\leq$ num\_edges $\leq$ 10 and 2 $\leq$ depth $\leq$ 5, and the following table reports the average accuracy varying depth.

\begin{table}[]
\centering
\small
\begin{tabular}{c|cccc}
Depth & 2 & 3 & 4 & 5 \\
\midrule
\makecell{QwQ-32B}     & 11.11 \%    & 32.5 \% & 37.14 \% & 33.33 \%      \\
\end{tabular}
\caption{Trace generation accuracy using a reasoning model (QwQ-32B).}
\label{tab:reasoning_accuracy}
\end{table}

Considering the accuracy of \sys in~\autoref{fig:accuracy_by_depth} is higher than 75\% for all depth parameters, the reasoning model with few-shot settings does not show comparable results even with easier settings (e.g., num\_edges $\leq$ 10). The results suggest that their reasoning mechanisms may not effectively capture the nuances of trace synthesis (e.g., counting depth and the number of edges), highlighting the need for approaches tailored to the specific task to generate traces.

\subsection{TraceLLM for Batch Job Traces}
\label{appendix:batch_job_gen}

\begin{figure}[]
    \centering
    \includegraphics[width=0.8\linewidth]{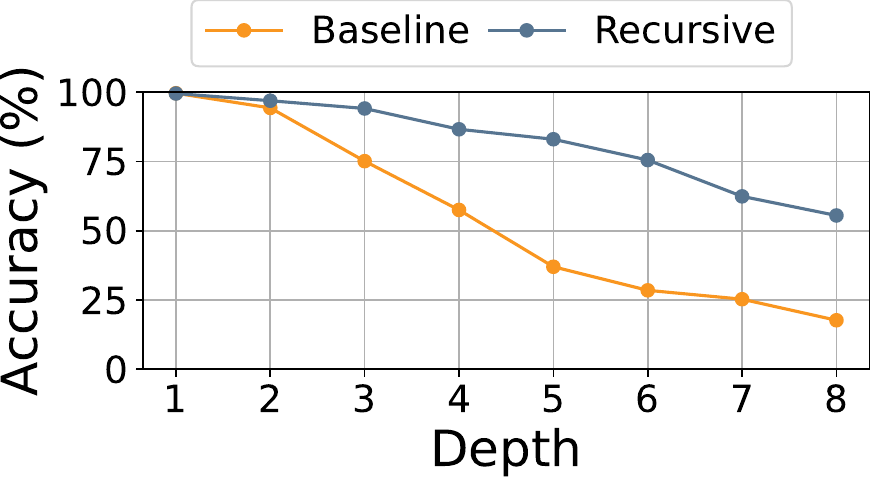}
     \caption{Accuracy to generate valid batch job traces.}
     \label{fig:batch_job_accuracy}
\end{figure}

While we study microservice call graphs in our paper, our method is not limited to microservice traces as stated in the limitations section. Our approach is designed to model structural dependencies within service interactions, which are fundamental properties not exclusive to microservices. The hierarchical nature of our method allows it to adapt to different levels of abstraction in distributed systems, making it applicable to other service architectures, such as batch job requests and serverless workflows.

We conduct additional experiments on generating batch job traces, a type of trace with hierarchical structures. A batch job is represented as a Directed Acyclic Graph (DAG) of tasks, where each task has attributes such as required CPU/memory resources, response time, and the number of instances. Tasks within a DAG have dependencies on one another. We measure accuracy by verifying that the synthetic traces form connected DAGs and satisfy constraints similar to those described in~\autoref{appendix:validation_rules}, such as the relationships between start and finish times.

Using batch job traces from the Alibaba 2018 cluster dataset~\citep{batchjobalibaba}, we train a Llama 3.2 3B model on 0.5B tokens and evaluate structured reasoning accuracy (Section 4.1) for both GReaT and our recursive method. In~\autoref{fig:batch_job_accuracy}, we report accuracy across different depths and task counts in prompts, with 1000 cases per depth configuration. As shown in the table below, the baseline model's accuracy declines rapidly as depth increases. In contrast, our recursive method maintains significantly higher accuracy, with up to a 47-percentage-point gap over the baseline, demonstrating its superior ability to handle complex hierarchical structures. We anticipate further improvements with instruction-tuning.

%%%%%%%%%%%%%%%%%%%%%%%%%%%%%%%%%%%%%%%%%%%%%%%%%%%%%%%%%%%%%%%%%%%%%%%%%%%%%%%
%%%%%%%%%%%%%%%%%%%%%%%%%%%%%%%%%%%%%%%%%%%%%%%%%%%%%%%%%%%%%%%%%%%%%%%%%%%%%%%

%% file: relatedwork.tex
\section{Other Related Work}\label{related}
% \aditya{this seems redundant given the earlier discussion, especially the first para? Start this section "Our work in the overall context of related efforts..." and shrink these texts}

% Our work in the overall context of related efforts in using LLMs to 
\mypara{Adapting LLMs for Specific Domains}
Pre-trained LLMs are increasingly adapted for specialized domains due to their vast, diverse training datasets, which enable broad generalization capabilities. Examples include fine-tuning LLMs for programming \citep{roziere2023code}, quantitative reasoning \citep{minerva}, and semiconductor manufacturing \citep{chipnemo}. Our work is the first to apply this approach to computer system traces involving data with specific structures and constraints. Our focus is on generating synthetic trace data by fine-tuning these models to handle the specific requirements of this domain.

\mypara{Making Language Models Follow Instructions}
Recent advancements have focused on enhancing LLMs' ability to follow instructions through prompting \citep{li-liang-2021-prefix, shin2020autoprompt, wei2022chain} and instruction tuning \citep{Ouyang2022TrainingLM, wei2021finetuned, flan}. These two sets of methods are relevant to our setting since they augment powerful pre-trained LLMs to improve their performance on new tasks. Our approach seeks to refine output expressiveness within set prompts, aiming for greater fidelity in synthetic data production.

\mypara{Multi-step Reasoning with LLMs}
Iterating with LLMs over multiple steps is an effective strategy to solve complex problems. 
For instance, Tree-of-thoughts~\citep{yao2024tree} solves problems by decomposing into smaller thoughts and exploring diverse reasoning paths over different thoughts. 
% \cite{paranjape2023art, trivedi2022interleaving} propose to interact with external knowledge sources during LLM inferences.
Multi-step reasoning is also useful to handle long-context scenarios by summarizing iteratively~\citep{wang2023recursively} and diving into subproblems~\citep{lee2023recursion}.
In contrast to the above approaches, our approach learns to generate traces with specific structures and instructions for subsequent layers.

%% file: acl_latex.bbl
\begin{thebibliography}{50}
\providecommand{\natexlab}[1]{#1}

\bibitem[{Amazon()}]{amazon}
Amazon.
\newblock Aws whitepaper: Implementing microservices on aws.
\newblock \url{https://docs.aws.amazon.com/whitepapers/latest/microservices-on-aws/microservices-on-aws.html}.
\newblock Accessed: 2025-02-15.

\bibitem[{Bergsma et~al.(2021)Bergsma, Zeyl, Senderovich, and Beck}]{vm_trace}
Shane Bergsma, Timothy Zeyl, Arik Senderovich, and J.~Christopher Beck. 2021.
\newblock \href {https://doi.org/10.1145/3477132.3483590} {Generating complex, realistic cloud workloads using recurrent neural networks}.
\newblock In \emph{Proceedings of the ACM SIGOPS 28th Symposium on Operating Systems Principles}, SOSP '21, page 376–391, New York, NY, USA. Association for Computing Machinery.

\bibitem[{Borisov et~al.(2023)Borisov, Sessler, Leemann, Pawelczyk, and Kasneci}]{borisov2023language}
Vadim Borisov, Kathrin Sessler, Tobias Leemann, Martin Pawelczyk, and Gjergji Kasneci. 2023.
\newblock Language models are realistic tabular data generators.
\newblock In \emph{The Eleventh International Conference on Learning Representations}.

\bibitem[{Brown et~al.(2020)Brown, Mann, Ryder, Subbiah, Kaplan, Dhariwal, Neelakantan, Shyam, Sastry, Askell, Agarwal, Herbert-Voss, Krueger, Henighan, Child, Ramesh, Ziegler, Wu, Winter, Hesse, Chen, Sigler, Litwin, Gray, Chess, Clark, Berner, McCandlish, Radford, Sutskever, and Amodei}]{gpt3}
Tom Brown, Benjamin Mann, Nick Ryder, Melanie Subbiah, Jared~D Kaplan, Prafulla Dhariwal, Arvind Neelakantan, Pranav Shyam, Girish Sastry, Amanda Askell, Sandhini Agarwal, Ariel Herbert-Voss, Gretchen Krueger, Tom Henighan, Rewon Child, Aditya Ramesh, Daniel Ziegler, Jeffrey Wu, Clemens Winter, Chris Hesse, Mark Chen, Eric Sigler, Mateusz Litwin, Scott Gray, Benjamin Chess, Jack Clark, Christopher Berner, Sam McCandlish, Alec Radford, Ilya Sutskever, and Dario Amodei. 2020.
\newblock Language models are few-shot learners.
\newblock In \emph{Advances in Neural Information Processing Systems}, volume~33, pages 1877--1901. Curran Associates, Inc.

\bibitem[{Chung et~al.(2022)Chung, Hou, Longpre, Zoph, Tay, Fedus, Li, Wang, Dehghani, Brahma et~al.}]{flan}
Hyung~Won Chung, Le~Hou, Shayne Longpre, Barret Zoph, Yi~Tay, William Fedus, Eric Li, Xuezhi Wang, Mostafa Dehghani, Siddhartha Brahma, et~al. 2022.
\newblock Scaling instruction-finetuned language models.
\newblock \emph{arXiv preprint arXiv:2210.11416}.

\bibitem[{Dubey et~al.(2024)Dubey, Jauhri, Pandey, Kadian, Al-Dahle, Letman, Mathur, Schelten, Yang, Fan et~al.}]{dubey2024llama}
Abhimanyu Dubey, Abhinav Jauhri, Abhinav Pandey, Abhishek Kadian, Ahmad Al-Dahle, Aiesha Letman, Akhil Mathur, Alan Schelten, Amy Yang, Angela Fan, et~al. 2024.
\newblock The llama 3 herd of models.
\newblock \emph{arXiv preprint arXiv:2407.21783}.

\bibitem[{Gan et~al.(2019)Gan, Zhang, Cheng, Shetty, Rathi, Katarki, Bruno, Hu, Ritchken, Jackson, Hu, Pancholi, He, Clancy, Colen, Wen, Leung, Wang, Zaruvinsky, Espinosa, Lin, Liu, Padilla, and Delimitrou}]{deathstarbench}
Yu~Gan, Yanqi Zhang, Dailun Cheng, Ankitha Shetty, Priyal Rathi, Nayan Katarki, Ariana Bruno, Justin Hu, Brian Ritchken, Brendon Jackson, Kelvin Hu, Meghna Pancholi, Yuan He, Brett Clancy, Chris Colen, Fukang Wen, Catherine Leung, Siyuan Wang, Leon Zaruvinsky, Mateo Espinosa, Rick Lin, Zhongling Liu, Jake Padilla, and Christina Delimitrou. 2019.
\newblock \href {https://doi.org/10.1145/3297858.3304013} {An open-source benchmark suite for microservices and their hardware-software implications for cloud \& edge systems}.
\newblock In \emph{Proceedings of the Twenty-Fourth International Conference on Architectural Support for Programming Languages and Operating Systems}, ASPLOS '19, page 3–18, New York, NY, USA. Association for Computing Machinery.

\bibitem[{Goodfellow et~al.(2014)Goodfellow, Pouget-Abadie, Mirza, Xu, Warde-Farley, Ozair, Courville, and Bengio}]{gan}
Ian Goodfellow, Jean Pouget-Abadie, Mehdi Mirza, Bing Xu, David Warde-Farley, Sherjil Ozair, Aaron Courville, and Yoshua Bengio. 2014.
\newblock Generative adversarial nets.
\newblock \emph{Advances in neural information processing systems}, 27.

\bibitem[{Gunasekar et~al.(2023)Gunasekar, Zhang, Aneja, Mendes, Del~Giorno, Gopi, Javaheripi, Kauffmann, de~Rosa, Saarikivi et~al.}]{phi}
Suriya Gunasekar, Yi~Zhang, Jyoti Aneja, Caio C{\'e}sar~Teodoro Mendes, Allie Del~Giorno, Sivakanth Gopi, Mojan Javaheripi, Piero Kauffmann, Gustavo de~Rosa, Olli Saarikivi, et~al. 2023.
\newblock Textbooks are all you need.
\newblock \emph{arXiv preprint arXiv:2306.11644}.

\bibitem[{Guo et~al.(2019)Guo, Chang, Wang, Ding, Feng, Mao, and Bao}]{batchjobalibaba}
Jing Guo, Zihao Chang, Sa~Wang, Haiyang Ding, Yihui Feng, Liang Mao, and Yungang Bao. 2019.
\newblock \href {https://doi.org/10.1145/3326285.3329074} {Who limits the resource efficiency of my datacenter: an analysis of alibaba datacenter traces}.
\newblock In \emph{Proceedings of the International Symposium on Quality of Service}, IWQoS '19, New York, NY, USA. Association for Computing Machinery.

\bibitem[{Ho et~al.(2020)Ho, Jain, and Abbeel}]{diffusion}
Jonathan Ho, Ajay Jain, and Pieter Abbeel. 2020.
\newblock Denoising diffusion probabilistic models.
\newblock \emph{Advances in neural information processing systems}, 33:6840--6851.

\bibitem[{Hu et~al.(2022)Hu, Shen, Wallis, Allen-Zhu, Li, Wang, Wang, and Chen}]{hu2022lora}
Edward~J Hu, Yelong Shen, Phillip Wallis, Zeyuan Allen-Zhu, Yuanzhi Li, Shean Wang, Lu~Wang, and Weizhu Chen. 2022.
\newblock Lo{RA}: Low-rank adaptation of large language models.
\newblock In \emph{International Conference on Learning Representations}.

\bibitem[{Huye et~al.(2024)Huye, Liu, and Sambasivan}]{alibaba_inconsistency}
Darby Huye, Lan Liu, and Raja~R. Sambasivan. 2024.
\newblock \href {https://doi.org/10.1145/3629526.3645043} {Systemizing and mitigating topological inconsistencies in alibaba's microservice call-graph datasets}.
\newblock In \emph{Proceedings of the 15th ACM/SPEC International Conference on Performance Engineering}, ICPE '24, page 276–285, New York, NY, USA. Association for Computing Machinery.

\bibitem[{Huye et~al.(2023)Huye, Shkuro, and Sambasivan}]{meta_microservice}
Darby Huye, Yuri Shkuro, and Raja~R. Sambasivan. 2023.
\newblock Lifting the veil on {Meta{\textquoteright}s} microservice architecture: Analyses of topology and request workflows.
\newblock In \emph{2023 USENIX Annual Technical Conference (USENIX ATC 23)}, pages 419--432, Boston, MA. USENIX Association.

\bibitem[{Ikram et~al.(2022)Ikram, Chakraborty, Mitra, Saini, Bagchi, and Kocaoglu}]{rcd}
Azam Ikram, Sarthak Chakraborty, Subrata Mitra, Shiv Saini, Saurabh Bagchi, and Murat Kocaoglu. 2022.
\newblock Root cause analysis of failures in microservices through causal discovery.
\newblock In \emph{Advances in Neural Information Processing Systems}, volume~35, pages 31158--31170. Curran Associates, Inc.

\bibitem[{Jiang et~al.(2023)Jiang, Liu, Gember-Jacobson, Schmitt, Bronzino, and Feamster}]{netfusion}
Xi~Jiang, Shinan Liu, Aaron Gember-Jacobson, Paul Schmitt, Francesco Bronzino, and Nick Feamster. 2023.
\newblock \href {https://doi.org/10.1145/3626111.3628196} {Generative, high-fidelity network traces}.
\newblock In \emph{Proceedings of the 22nd ACM Workshop on Hot Topics in Networks}, HotNets '23, page 131–138, New York, NY, USA. Association for Computing Machinery.

\bibitem[{Kingma and Welling(2013)}]{vae}
Diederik~P Kingma and Max Welling. 2013.
\newblock Auto-encoding variational bayes.
\newblock \emph{arXiv preprint arXiv:1312.6114}.

\bibitem[{Lee and Kim(2023)}]{lee2023recursion}
Soochan Lee and Gunhee Kim. 2023.
\newblock Recursion of thought: A divide-and-conquer approach to multi-context reasoning with language models.
\newblock \emph{arXiv preprint arXiv:2306.06891}.

\bibitem[{Lewkowycz et~al.(2022)Lewkowycz, Andreassen, Dohan, Dyer, Michalewski, Ramasesh, Slone, Anil, Schlag, Gutman-Solo et~al.}]{minerva}
Aitor Lewkowycz, Anders Andreassen, David Dohan, Ethan Dyer, Henryk Michalewski, Vinay Ramasesh, Ambrose Slone, Cem Anil, Imanol Schlag, Theo Gutman-Solo, et~al. 2022.
\newblock Solving quantitative reasoning problems with language models.
\newblock \emph{Advances in Neural Information Processing Systems}, 35:3843--3857.

\bibitem[{Li et~al.(2024)Li, Chen, Chen, He, Gu, and Zhou}]{li2024selective}
Ming Li, Lichang Chen, Jiuhai Chen, Shwai He, Jiuxiang Gu, and Tianyi Zhou. 2024.
\newblock Selective reflection-tuning: Student-selected data recycling for llm instruction-tuning.
\newblock \emph{arXiv preprint arXiv:2402.10110}.

\bibitem[{Li and Liang(2021)}]{li-liang-2021-prefix}
Xiang~Lisa Li and Percy Liang. 2021.
\newblock \href {https://doi.org/10.18653/v1/2021.acl-long.353} {Prefix-tuning: Optimizing continuous prompts for generation}.
\newblock In \emph{Proceedings of the 59th Annual Meeting of the Association for Computational Linguistics and the 11th International Joint Conference on Natural Language Processing (Volume 1: Long Papers)}, pages 4582--4597, Online. Association for Computational Linguistics.

\bibitem[{Lin et~al.(2020)Lin, Jain, Wang, Fanti, and Sekar}]{doppelganger}
Zinan Lin, Alankar Jain, Chen Wang, Giulia Fanti, and Vyas Sekar. 2020.
\newblock Using gans for sharing networked time series data: Challenges, initial promise, and open questions.
\newblock In \emph{Proceedings of the ACM Internet Measurement Conference}, pages 464--483.

\bibitem[{Liu et~al.(2024)Liu, Li, Wu, and Lee}]{llava}
Haotian Liu, Chunyuan Li, Qingyang Wu, and Yong~Jae Lee. 2024.
\newblock Visual instruction tuning.
\newblock \emph{Advances in neural information processing systems}, 36.

\bibitem[{Liu et~al.(2023)Liu, Ene, Kirby, Cheng, Pinckney, Liang, Alben, Anand, Banerjee, Bayraktaroglu et~al.}]{chipnemo}
Mingjie Liu, Teodor-Dumitru Ene, Robert Kirby, Chris Cheng, Nathaniel Pinckney, Rongjian Liang, Jonah Alben, Himyanshu Anand, Sanmitra Banerjee, Ismet Bayraktaroglu, et~al. 2023.
\newblock Chipnemo: Domain-adapted llms for chip design.
\newblock \emph{arXiv preprint arXiv:2311.00176}.

\bibitem[{Loshchilov and Hutter(2017)}]{adamw}
Ilya Loshchilov and Frank Hutter. 2017.
\newblock Decoupled weight decay regularization.
\newblock \emph{arXiv preprint arXiv:1711.05101}.

\bibitem[{Luo et~al.(2021)Luo, Xu, Lu, Ye, Xu, Zhang, Ding, He, and Xu}]{alibaba_microservice_v2021}
Shutian Luo, Huanle Xu, Chengzhi Lu, Kejiang Ye, Guoyao Xu, Liping Zhang, Yu~Ding, Jian He, and Chengzhong Xu. 2021.
\newblock \href {https://doi.org/10.1145/3472883.3487003} {Characterizing microservice dependency and performance: Alibaba trace analysis}.
\newblock In \emph{Proceedings of the ACM Symposium on Cloud Computing}, SoCC '21, page 412–426, New York, NY, USA. Association for Computing Machinery.

\bibitem[{Luo et~al.(2022)Luo, Xu, Ye, Xu, Zhang, Yang, and Xu}]{madu}
Shutian Luo, Huanle Xu, Kejiang Ye, Guoyao Xu, Liping Zhang, Guodong Yang, and Chengzhong Xu. 2022.
\newblock \href {https://doi.org/10.1145/3542929.3563477} {The power of prediction: microservice auto scaling via workload learning}.
\newblock In \emph{Proceedings of the 13th Symposium on Cloud Computing}, SoCC '22, page 355–369, New York, NY, USA. Association for Computing Machinery.

\bibitem[{Netflix()}]{netflix}
Netflix.
\newblock Netflix tech blog: Microservices.
\newblock \url{https://netflixtechblog.com/tagged/microservices}.
\newblock Accessed: 2025-02-15.

\bibitem[{Nye et~al.(2021)Nye, Andreassen, Gur-Ari, Michalewski, Austin, Bieber, Dohan, Lewkowycz, Bosma, Luan, Sutton, and Odena}]{51142}
Maxwell Nye, Anders Andreassen, Guy Gur-Ari, Henryk~Witold Michalewski, Jacob Austin, David Bieber, David~Martin Dohan, Aitor Lewkowycz, Maarten~Paul Bosma, David Luan, Charles Sutton, and Augustus Odena. 2021.
\newblock Show your work: Scratchpads for intermediate computation with language models.
\newblock Https://arxiv.org/abs/2112.00114.

\bibitem[{Ouyang et~al.(2022)Ouyang, Wu, Jiang, Almeida, Wainwright, Mishkin, Zhang, Agarwal, Slama, Ray, Schulman, Hilton, Kelton, Miller, Simens, Askell, Welinder, Christiano, Leike, and Lowe}]{Ouyang2022TrainingLM}
Long Ouyang, Jeff Wu, Xu~Jiang, Diogo Almeida, Carroll~L. Wainwright, Pamela Mishkin, Chong Zhang, Sandhini Agarwal, Katarina Slama, Alex Ray, John Schulman, Jacob Hilton, Fraser Kelton, Luke~E. Miller, Maddie Simens, Amanda Askell, Peter Welinder, Paul~Francis Christiano, Jan Leike, and Ryan~J. Lowe. 2022.
\newblock Training language models to follow instructions with human feedback.
\newblock \emph{ArXiv}, abs/2203.02155.

\bibitem[{Qiu et~al.(2020)Qiu, Banerjee, Jha, Kalbarczyk, and Iyer}]{firm}
Haoran Qiu, Subho~S. Banerjee, Saurabh Jha, Zbigniew~T. Kalbarczyk, and Ravishankar~K. Iyer. 2020.
\newblock {FIRM}: An intelligent fine-grained resource management framework for {SLO-Oriented} microservices.
\newblock In \emph{14th USENIX Symposium on Operating Systems Design and Implementation (OSDI 20)}, pages 805--825. USENIX Association.

\bibitem[{Roziere et~al.(2023)Roziere, Gehring, Gloeckle, Sootla, Gat, Tan, Adi, Liu, Remez, Rapin et~al.}]{roziere2023code}
Baptiste Roziere, Jonas Gehring, Fabian Gloeckle, Sten Sootla, Itai Gat, Xiaoqing~Ellen Tan, Yossi Adi, Jingyu Liu, Tal Remez, J{\'e}r{\'e}my Rapin, et~al. 2023.
\newblock Code llama: Open foundation models for code.
\newblock \emph{arXiv preprint arXiv:2308.12950}.

\bibitem[{Sanh et~al.(2021)Sanh, Webson, Raffel, Bach, Sutawika, Alyafeai, Chaffin, Stiegler, Scao, Raja et~al.}]{t_zero}
Victor Sanh, Albert Webson, Colin Raffel, Stephen~H Bach, Lintang Sutawika, Zaid Alyafeai, Antoine Chaffin, Arnaud Stiegler, Teven~Le Scao, Arun Raja, et~al. 2021.
\newblock Multitask prompted training enables zero-shot task generalization.
\newblock \emph{arXiv preprint arXiv:2110.08207}.

\bibitem[{Shen et~al.(2024)Shen, Tenenholtz, Hall, Alvarez-Melis, and Fusi}]{tag-llm}
Junhong Shen, Neil Tenenholtz, James~Brian Hall, David Alvarez-Melis, and Nicolo Fusi. 2024.
\newblock \href {https://arxiv.org/abs/2402.05140} {Tag-llm: Repurposing general-purpose llms for specialized domains}.
\newblock \emph{Preprint}, arXiv:2402.05140.

\bibitem[{Sherstinsky(2020)}]{rnn}
Alex Sherstinsky. 2020.
\newblock Fundamentals of recurrent neural network (rnn) and long short-term memory (lstm) network.
\newblock \emph{Physica D: Nonlinear Phenomena}, 404:132306.

\bibitem[{Shin et~al.(2020)Shin, Razeghi, Logan~IV, Wallace, and Singh}]{shin2020autoprompt}
Taylor Shin, Yasaman Razeghi, Robert~L Logan~IV, Eric Wallace, and Sameer Singh. 2020.
\newblock Autoprompt: Eliciting knowledge from language models with automatically generated prompts.
\newblock \emph{arXiv preprint arXiv:2010.15980}.

\bibitem[{Singhvi et~al.(2021)Singhvi, Balasubramanian, Houck, Shaikh, Venkataraman, and Akella}]{atoll}
Arjun Singhvi, Arjun Balasubramanian, Kevin Houck, Mohammed~Danish Shaikh, Shivaram Venkataraman, and Aditya Akella. 2021.
\newblock Atoll: A scalable low-latency serverless platform.
\newblock In \emph{Proceedings of the ACM Symposium on Cloud Computing}, pages 138--152.

\bibitem[{Team et~al.(2025)Team, Bai, Bao, Chen, Chen, Chen, Chen, Chen, Chen, Chen et~al.}]{kimik2}
Kimi Team, Yifan Bai, Yiping Bao, Guanduo Chen, Jiahao Chen, Ningxin Chen, Ruijue Chen, Yanru Chen, Yuankun Chen, Yutian Chen, et~al. 2025.
\newblock Kimi k2: Open agentic intelligence.
\newblock \emph{arXiv preprint arXiv:2507.20534}.

\bibitem[{Team(2025)}]{qwq32b}
Qwen Team. 2025.
\newblock \href {https://qwenlm.github.io/blog/qwq-32b/} {Qwq-32b: Embracing the power of reinforcement learning}.

\bibitem[{Touvron et~al.(2023)Touvron, Lavril, Izacard, Martinet, Lachaux, Lacroix, Rozi{\`e}re, Goyal, Hambro, Azhar et~al.}]{touvron2023llama}
Hugo Touvron, Thibaut Lavril, Gautier Izacard, Xavier Martinet, Marie-Anne Lachaux, Timoth{\'e}e Lacroix, Baptiste Rozi{\`e}re, Naman Goyal, Eric Hambro, Faisal Azhar, et~al. 2023.
\newblock Llama: Open and efficient foundation language models.
\newblock \emph{arXiv preprint arXiv:2302.13971}.

\bibitem[{Uber()}]{uber}
Uber.
\newblock Introducing domain-oriented microservice architecture.
\newblock \url{https://www.uber.com/blog/microservice-architecture/}.
\newblock Accessed: 2025-02-15.

\bibitem[{Wang et~al.(2023)Wang, Ding, Cao, Tian, Wang, Tao, and Guo}]{wang2023recursively}
Qingyue Wang, Liang Ding, Yanan Cao, Zhiliang Tian, Shi Wang, Dacheng Tao, and Li~Guo. 2023.
\newblock Recursively summarizing enables long-term dialogue memory in large language models.
\newblock \emph{arXiv preprint arXiv:2308.15022}.

\bibitem[{Wei et~al.(2021)Wei, Bosma, Zhao, Guu, Yu, Lester, Du, Dai, and Le}]{wei2021finetuned}
Jason Wei, Maarten Bosma, Vincent~Y Zhao, Kelvin Guu, Adams~Wei Yu, Brian Lester, Nan Du, Andrew~M Dai, and Quoc~V Le. 2021.
\newblock Finetuned language models are zero-shot learners.
\newblock \emph{arXiv preprint arXiv:2109.01652}.

\bibitem[{Wei et~al.(2022)Wei, Wang, Schuurmans, Bosma, Xia, Chi, Le, Zhou et~al.}]{wei2022chain}
Jason Wei, Xuezhi Wang, Dale Schuurmans, Maarten Bosma, Fei Xia, Ed~Chi, Quoc~V Le, Denny Zhou, et~al. 2022.
\newblock Chain-of-thought prompting elicits reasoning in large language models.
\newblock \emph{Advances in Neural Information Processing Systems}, 35:24824--24837.

\bibitem[{Xie et~al.(2023)Xie, Xu, Chen, Li, Jiang, Su, Wang, and Pei}]{tracevae}
Zhe Xie, Haowen Xu, Wenxiao Chen, Wanxue Li, Huai Jiang, Liangfei Su, Hanzhang Wang, and Dan Pei. 2023.
\newblock \href {https://doi.org/10.1145/3543507.3583215} {Unsupervised anomaly detection on microservice traces through graph vae}.
\newblock In \emph{Proceedings of the ACM Web Conference 2023}, WWW '23, page 2874–2884, New York, NY, USA. Association for Computing Machinery.

\bibitem[{Xu et~al.(2019)Xu, Skoularidou, Cuesta-Infante, and Veeramachaneni}]{ctgan}
Lei Xu, Maria Skoularidou, Alfredo Cuesta-Infante, and Kalyan Veeramachaneni. 2019.
\newblock Modeling tabular data using conditional gan.
\newblock \emph{Advances in neural information processing systems}, 32.

\bibitem[{Yao et~al.(2024)Yao, Yu, Zhao, Shafran, Griffiths, Cao, and Narasimhan}]{yao2024tree}
Shunyu Yao, Dian Yu, Jeffrey Zhao, Izhak Shafran, Tom Griffiths, Yuan Cao, and Karthik Narasimhan. 2024.
\newblock Tree of thoughts: Deliberate problem solving with large language models.
\newblock \emph{Advances in Neural Information Processing Systems}, 36.

\bibitem[{Yin et~al.(2022)Yin, Lin, Jin, Fanti, and Sekar}]{netshare}
Yucheng Yin, Zinan Lin, Minhao Jin, Giulia Fanti, and Vyas Sekar. 2022.
\newblock \href {https://doi.org/10.1145/3544216.3544251} {{Practical GAN-based synthetic IP header trace generation using NetShare}}.
\newblock In \emph{Proceedings of the ACM SIGCOMM 2022 Conference}, SIGCOMM '22, page 458–472, New York, NY, USA. Association for Computing Machinery.

\bibitem[{Zhang et~al.(2024)Zhang, Zhou, Elnikety, and Delimitrou}]{ursa}
Y.~Zhang, Z.~Zhou, S.~Elnikety, and C.~Delimitrou. 2024.
\newblock \href {https://doi.org/10.1109/HPCA57654.2024.00077} {{Ursa: Lightweight Resource Management for Cloud-Native Microservices}}.
\newblock In \emph{2024 IEEE International Symposium on High-Performance Computer Architecture (HPCA)}, pages 954--969, Los Alamitos, CA, USA. IEEE Computer Society.

\bibitem[{Zhang et~al.(2022)Zhang, Ramanathan, Raj, Parwal, Sherwood, and Chabbi}]{crisp}
Zhizhou Zhang, Murali~Krishna Ramanathan, Prithvi Raj, Abhishek Parwal, Timothy Sherwood, and Milind Chabbi. 2022.
\newblock {CRISP}: Critical path analysis of {Large-Scale} microservice architectures.
\newblock In \emph{2022 USENIX Annual Technical Conference (USENIX ATC 22)}, pages 655--672, Carlsbad, CA. USENIX Association.

\end{thebibliography}
